\documentclass[review,sort&compress]{elsarticle}

\usepackage[utf8]{inputenc}
\usepackage[english]{babel}
\usepackage{graphicx}
\usepackage{dcolumn}
\usepackage{bm}
\usepackage{amsmath}
\usepackage{amsfonts}

\usepackage{xcolor}
\usepackage{microtype}
\usepackage{ragged2e}

\usepackage[colorlinks,allcolors=blue]{hyperref}

\newcommand{\mc}{\mathcal}

\DeclareMathOperator{\Ln}{Ln}

\DeclareMathOperator{\Grad}{Grad}

\begin{document}

\title{A unified framework for divergences, free energies, and Fokker-Planck equations}

\author[1,2]{Anna L. F. Lucchi}
\ead{annalauraferreiralucchi@gmail.com}

\author[1,2]{Jean H. Y. Passos}
\ead{jhypassos@gmail.com}

\author[3]{Max Jauregui}
\ead{max.rodriguez@ufmt.br}

\author[1,2]{Renio S. Mendes}
\ead{rsmendes@dfi.uem.br}

\affiliation[1]{organization={Departamento de Física, Universidade Estadual de Maringá}, addressline={Av. Colombo, 5790, 87020-900}, city={Maringá, PR}, country={Brazil}}

\affiliation[2]{organization={National Institute of Science and Technology for Complex Systems}, addressline={Rua Xavier Sigaud, 150, 22290-180}, city={Rio de Janeiro, RJ}, country={Brazil}}

\affiliation[3]{organization={Faculdade de Ciência e Tecnologia, Universidade Federal de Mato Grosso}, addressline={Av. Fernando Corrêa da Costa, 2367, Bairro Boa Esperança, 78060-900}, city={Cuiabá, MT}, country={Brazil}}

\date{\today}

\begin{abstract}
Many efforts have been made to explore systems that show significant deviations from predictions related to the standard statistical mechanics. The present work introduces a unified formalism that connects divergences, generalized free energies, generalized Fokker-Planck equations, and $H$-theorem. This framework is applied here in a range of scenarios, illustrating both established and novel results. In many cases, the approach begins with a free energy functional that explicitly includes a potential energy term, leading to a direct relation between this energy and the stationary solution. Conversely, when a divergence is used as free energy, the associated Fokker-Planck-like equation lacks any explicit dependence on the potential energy, depending instead on the stationary solution. To restore a potential-based interpretation, an additional relation between the stationary solution and the potential energy must be imposed. This duality underlines the flexibility of the formalism and its capacity to adapt to systems where the potential energy is unknown or unnecessary.
\end{abstract}


\maketitle

\section{Introduction}\label{introduction}

In the last four decades, there
has been a growing interest in investigating situations related to deviations from the formalism of Boltzmann-Gibbs-Shannon statistical mechanics. To deal with these cases, some proposals, such as the Tsallis~\cite{Tsallis1988} and Kaniadakis~\cite{kaniadakis2001non} entropies, have been considered. In the Tsallis context, the current investigations include dynamics of chaotic maps~\cite{tirnakli2016standard}, tumor edge detection~\cite{tariq2021tumor}, Raman scattering~\cite{gupta2022stimulated}, high-energy physics~\cite{makke2024data}, and chemical reactions~\cite{PhysRevResearch.7.L012081}, among a myriad of others~\cite{tsallis2009introduction}. Concerning the Kaniadakis approach, recent advances are related, for instance, to dark energy~\cite{drepanou2022kaniadakis}, electron-acoustic waves~\cite{irshad2023effect}, nuclear reactors~\cite{e25030478}, and relativity~\cite{e26050406}. A comprehensive list of investigations involving the Kaniadakis entropy can be found in~\cite{e27030247}. Moreover, even a formulation of statistical mechanics based on the Rényi entropy has also been investigated~\cite{LENZI2000337}. 

In the Tsallis and Kaniadakis contexts, time-dependent situations involving (nonlinear) Fokker-Planck equations were also investigated~\cite{plastino1995non, PhysRevE.54.R2197, hirica2022lie}. Their stationary solutions correspond to the canonical distributions and there are $H$-theorems related to them. A few further but representative developments based on (nonlinear) Fokker-Planck equations and their $H$-theorems begin with~\cite{shiino2001free, kaniadakis2001h, schwammle2007consequences, schwammle2007general}, while extensions of these studies include inhomogeneous cases~\cite{sicuro2016nonlinear, jauregui2021stationary}, fractional derivatives~\cite{SILVA200765, lenzi2024generalized}, coupled nonlinear equations~\cite{plastino2022h, evangelista2023nonlinear}, and measures of complexity~\cite{plastino2024evolution}. In addition, several applications and methodological approaches to nonlinear Fokker-Planck equations can be found in~\cite{frank2005nonlinear}. From a broader perspective and a formal point of view, we highlight that some general aspects are independent of the particular thermostatistics considered. For instance, this is observed in the universality of the Legendre transform structure, regardless of the entropy employed~\cite{plastino1997universality, mendes1997some}, and in the classification of entropies~\cite{hanel2011comprehensive}.

Commonly, entropies are associated with divergences which quantify how two probability distributions differ from each other~\cite{abu2019effects}. Terms such as relative entropy, information divergence, information gain, and even distance are also used to refer to divergences~\cite{deza2009encyclopedia}. A divergence between two probability density functions, $\rho$ and $\rho_0$, is here denoted by $\mathcal{D}(\rho,\rho_0)$ and satisfies the inequality $\mathcal{D}(\rho,\rho_0) \ge 0$ with $\mathcal{D}(\rho,\rho_0)=0$ if and only if $\rho=\rho_0$~\cite{deza2009encyclopedia}. The Shannon entropy is closely related to the well-known Kullback-Leibler divergence~\cite{KullbackLeibler1951}. Applications of this divergence are vast, including image retrieval and classification~\cite{sakji2010fast, cui2012statistical, qin2015region, cui2016comparison}, text characterization~\cite{dhillon2003divisive, huang2008similarity}, seismic information extraction~\cite{sun2020ml}, decision-making systems~\cite{ponti2017decision}, optimization problems~\cite{kobayashi2022optimistic}, and machine learning~\cite{clim2018kullback}, among many others. Analogously, there are the Rényi entropy and a related divergence~\cite{Rényi1961measures}, which have a wide range of applications~\cite{he2003generalized, li2019certified, mironov2019r, chowdhury2020Rényi, knoblauch2022optimization}. Similarly, the Tsallis entropy~\cite{Tsallis1988} and an associated divergence~\cite{Tsallisqdiv} have been employed in quantum physics~\cite{abe2003nonadditive, rastegin2016quantum}, variational inference~\cite{wang2021variational}, radioactive source detection~\cite{fu2022tsallis}, and Markov decision processes~\cite{leleux2024sparse}. Likewise, the Kaniadakis entropy~\cite{kaniadakis2001non} and a corresponding divergence~\cite{kaniadakisdivergence} have also been studied. All these entropies and divergences recover the Shannon entropy and the Kullback-Leibler divergence as limiting cases. More general divergences are the $f$-divergence~\cite{Rényi1961measures} (also called Csiszár~\cite{csiszar1974information}), the Bregman divergence~\cite{bregmandivergence} and the Burbea-Rao divergence~\cite{burbea1982convexity}.

As should be clear from the previous discussion, depending on the problem and the data, some aspects of an analysis may be more relevant than others; therefore, a triad composed of entropy, divergence, and the Fokker-Planck equation may be more appropriate than another one. By presenting a unified formalism, this work connects divergence, (generalized) free energy, the (generalized) Fokker-Planck equation, and the $H$-theorem, and demonstrates its versatility through both new and known applications. Our approach is consistent with previous works that connect the difference between two free energies with divergence in Shannon, Tsallis and Kaniadakis contexts~\cite{shiino1998h, qian2001relative, abe2005necessity, pennini2021free, hirica2022lie}. The present work is organized as follows. In Sec.~\ref{section1}, several entropies and their corresponding divergences are reported. We begin with the Shannon and Kullback-Leibler pair. The Tsallis and Kaniadakis scenarios are addressed through the introduction of the $q$-logarithm and $\kappa$-logarithm functions, respectively. Likewise, the Rényi case is considered. These scenarios are also extended to incorporate generalized logarithms. In addition, the Bregman and Burbea-Rao divergences are discussed. In Sec.~\ref{Divergences-as-free-energies}, we introduce a unified approach that identifies each divergence with the difference between two generalized free energies. This integrated perspective is achieved via a Fokker-Planck-like equation and its $H$-theorem. Given a free energy, a Fokker-Planck equation is obtained and its stationary solution is connected with the minimum of this free energy. In Sec.~\ref{Applications}, for all divergences presented in Sec.~\ref{section1}, the corresponding free energies are exhibited as well as their Fokker-Planck equations and stationary solutions. We show that some divergences (free energies) are connected with drift forces and, therefore, with their potential energies. When a stationary solution $\rho_0$ is not associated with a potential energy $V$, we consider the possibility of Fokker-Planck-like equations that explicitly depend on $\rho_0$, without directly involving $V$. The last section presents our concluding remarks.

\section{Entropies and divergences}\label{section1}

The Shannon entropy is a measure of the amount of information contained in a discrete probability distribution. This measure attains its maximum value, corresponding to the smallest quantity of information, when applied to the uniform distribution. For an absolutely continuous distribution with density $\rho$, the Shannon entropy is defined by~\cite{6773024}
\begin{equation}
\label{shannonentropy}
    S(\rho)= \int_{-\infty}^\infty \rho(x) \ln \left( \frac{1}{\rho(x)} \right) dx \, .
\end{equation}%
As pointed out by Shannon himself, the continuous version lacks some of the properties of the discrete case. For instance, it can be negative and it does not remain invariant under coordinate transformations. In contrast, the Kullback-Leibler divergence (sometimes called relative entropy) of two probability densities $\rho$ and 
$\rho_0$, formulated as~\cite{KullbackLeibler1951} 
\begin{equation}
\label{KLdiv}
    \mathcal{D}^{KL}(\rho , \rho_0) = -\int_{-\infty}^\infty \rho(x) \ln \left( \frac{\rho_0(x)}{\rho(x)} \right) dx \, , 
\end{equation}%
is always non-negative and remains invariant under coordinate transformations. The fact that $\mathcal{D}^{KL}(\rho,\rho_0)\ge 0$ follows directly from the inequality $\ln (x)\le x-1$, which holds for every $x>0$. Moreover, $\mathcal{D}^{KL}(\rho,\rho_0)=0$ if and only if $\rho(x)=\rho_0(x)$ almost everywhere, which means that the functions may differ on a set of points of measure zero~\cite{Durrett_2019}. This mathematical technicality will not be used in this work and consequently will be omitted in other similar situations. In the following of this section, we will present generalizations of the Shannon entropy and the Kullback-Leibler divergence.

\subsection{Tsallis entropy and divergence}

As reported in the introduction, a generalization of the Boltzmann-Gibbs statistical mechanics, based on the Tsallis entropy ($S_q^{T}$), has been applied in several studies. In the continuous case, this entropy can be written as~\cite{Tsallis1988}
\begin{equation}
\label{Sq}
    S_q^{T} (\rho)=  \int_{-\infty}^\infty \rho(x) \ln_q \left( \frac{1}{\rho(x)}\right) dx\,,
\end{equation}%
where the function $\ln_q(x)$, usually called $q$-logarithm, is given by~\cite{tsallis2009introduction}
\begin{equation}
\label{lnq}
    \ln_q (x) = \frac{x^{1-q}-1}{1-q}\qquad (q\ne 1)\, .
\end{equation}%
Since $\ln_q (x)\to \ln (x)$ as $q\to 1$, the entropy $S_q^{T}$ recovers the Shannon entropy when $q\to 1$. Replacing the logarithm with the $q$-logarithm in Eq.~(\ref{KLdiv}) leads to the following generalization of the Kullback-Leibler divergence~\cite{Tsallisqdiv}:
\begin{equation}
\label{qdiv}
\mathcal{D}_q^{T} (\rho, \rho_0) = -\int_{-\infty}^\infty \rho(x) \ln_q \left( \frac{\rho_0(x)}{\rho(x)} \right) dx \, .
\end{equation}%
For $q>0$, this divergence also satisfies the condition $\mathcal{D}_q^{T} (\rho , \rho_0) \ge 0$, where the equality holds if and only if $\rho(x) = \rho_0(x)$.

\subsection{Kaniadakis entropy and divergence}

Another generalized logarithm, proposed by Kaniadakis and usually referred to as the $\kappa$-logarithm, is provided by~\cite{kaniadakisentropy}
\begin{equation}
    \ln_{\kappa} (x)= \frac{x^{\kappa} - x^{-\kappa} }{2 \kappa}\qquad(\kappa\ne 0)\,,
\end{equation}%
hence $\ln_\kappa (x)\to \ln (x)$ when $\kappa\to 0$. By using the $\kappa$-logarithm, the Kaniadakis entropy~\cite{kaniadakisentropy} follows as
\begin{equation}
\label{kappa-div}
    S_{\kappa}^{K} (\rho) = \int_{-\infty}^\infty \rho(x) \ln_{\kappa} \left( \frac{1}{\rho(x)} \right) dx \, .
\end{equation}%
In connection with this entropy, it is natural to define the divergence~\cite{kaniadakisdivergence}
\begin{equation}
\label{divergencia-kaniadakis} \mathcal{D}_{\kappa}^{K} (\rho, \rho_0) = -\int_{-\infty}^\infty \rho(x) \ln_{\kappa} \left( \frac{\rho_0(x)}{\rho(x)} \right) dx \,.
\end{equation}%
If $|\kappa|\le 1$, this divergence satisfies the inequality $\mathcal{D}_k^{K}(\rho,\rho_0)\ge 0$, where the equality holds if, and only if, $\rho(x)=\rho_0(x)$.

\subsection{\texorpdfstring{$f$-divergence}{f-divergence}}

The usual logarithm, the $q$-logarithm and the $\kappa$-logarithm are particular cases of the generalized logarithm~\cite{naudts2004generalized}
\begin{equation}
\label{Lngeneralizado}
    \Ln (x) = \int_1^{x} \frac{dy}{g(y)}, 
\end{equation}%
where $g$ is a non-negative differentiable function such that $g(1)=1$ and $g'(y)>0$ for every $y>0$. Indeed, we recover the usual logarithm, the $q$-logarithm and the $\kappa$-logarithm by considering, respectively, $g(y)=y$, $g(y)=y^q$ and $g(y)=2y/(y^\kappa+y^{-\kappa})$. However, the condition $g'(y)>0$ leads to $q>0$ and $|\kappa| \leq 1$. As immediate consequences of Eq.~(\ref{Lngeneralizado}), we have that the generalized logarithm $\Ln(x)$ is monotonically increasing and concave on the interval $(0,\infty)$, and that $\Ln (1)=0$. Moreover, it can be proved that $\Ln (x)\le (x-1)/g(1)=x-1$ for every $x>0$.

The generalized logarithm $\Ln(x)$ enables us to define the generalized entropy
\begin{equation}
\label{Sgeneralizado}
     \mathcal{S}_{\Ln} (\rho)  =  \int_{-\infty}^{\infty} \rho(x) \, \Ln \left( \frac{1}{\rho(x)} \right) dx
\end{equation}%
and the corresponding divergence
\begin{equation}
\label{Dgeneralizado}
     \mathcal{D}_{\Ln} (\rho, \rho_0)  = - \int_{-\infty}^{\infty} \rho(x) \, \Ln \left( \frac{\rho_0(x)}{\rho(x)} \right) dx\,. 
\end{equation}%
Since  $\Ln(\rho_0 (x) / \rho (x)) \le \rho_0 (x) / \rho (x) -1$, it can be concluded that $\mathcal{D}_{\Ln}(\rho,\rho_0)\ge 0$, where the equality holds if and only if $\rho(x)=\rho_0(x)$.

A large number of divergences, including those given in Eqs.~(\ref{KLdiv}),~(\ref{qdiv}),~(\ref{kappa-div}) and~(\ref{Dgeneralizado}), belong to the family of $f$-divergences. In its most
general formulation, an f-divergence can be expressed as~\cite{Rényi1961measures, ali1966general, csiszar1972class,liese2006divergences}
\begin{equation}
\label{f-divergencia}
    \mathcal{D}_f (\rho, \rho_0)= \int_{-\infty}^\infty \rho_0(x) \, f \left( \frac{\rho(x)}{\rho_0(x)} \right)dx\,,
\end{equation}%
where $f$ is a function that is convex in the interval $(0,\infty)$ and satisfies $f(1)=0$, among other conditions that ensure the convergence of the integral. We note that if $f$ is a twice differentiable function, the divergences $\mathcal{D}_f(\rho,\rho_0)$ and $\mathcal{D}_{\Ln}(\rho,\rho_0)$ become equivalent when $f(x) = -C \, x\Ln(1/x) $, where $C$ is a positive constant.

\subsection{Rényi entropy and divergence}

A well-known example of a divergence that does not belong to the family of $f$-divergences is the Rényi one, which is defined by~\cite{Rényi1961measures}
\begin{equation}
\label{Renyidiv}
    \mathcal{D}_{\alpha}^{R} (\rho,\rho_0) = \frac{1}{\alpha-1} \ln \left( \int_{-\infty}^\infty \frac{[\rho(x)]^{\alpha}}{[\rho_0(x)]^{\alpha-1}} dx  \right) ,
\end{equation}%
with $\alpha>0$ and $\alpha\ne 1$. The form of this divergence is related to the Rényi entropy ~\cite{Rényi1961measures}
\begin{equation}
\label{Renyientropy}
    S_{\alpha}^{R} (\rho) = \frac{1}{1-\alpha} \ln \left( \int_{-\infty}^\infty [\rho(x)]^{\alpha} dx \right)\quad (\alpha>0 \, \,  \text{and} \, \, \alpha\ne 1)\,.
\end{equation}%
We can immediately verify that $\mathcal{D}_\alpha^{R}$ and $S_\alpha^{R}$ respectively recover the Kullback-Leibler divergence and the Shannon entropy in the limiting case $\alpha\to 1$. 

Equations~(\ref{Renyidiv}) and~(\ref{Renyientropy}) can also be extended by replacing the usual logarithm with a generalized one. In this direction, using the function $\Ln(x)$, defined in Eq.~(\ref{Lngeneralizado}), we propose the family of divergences \begin{equation}
\label{generalizedRenyi}
    \mathcal{D}_{\alpha, \Ln}^{R} (\rho, \rho_0) = \frac{1}{\alpha-1} \Ln \left( \int_{-\infty}^\infty \frac{[\rho(x)]^{\alpha}}{[\rho_0(x)]^{\alpha-1}} dx  \right)
\end{equation}%
and the family of entropies
\begin{equation}
\label{Renyiln}
     \mathcal{S}_{\alpha, \Ln}^{R} (\rho) = \frac{1}{1-\alpha} \Ln \left( \int_{-\infty}^\infty [\rho(x)]^{\alpha} dx \right) ,
\end{equation}%
with $\alpha>0$ and $\alpha \neq 1$. Remarkably, despite the arbitrariness in the definition of the function $ \Ln(x)$, Eqs.~(\ref{generalizedRenyi}) and~(\ref{Renyiln}) recover respectively the Kullback-Leibler divergence and the Shannon entropy in the limiting case $\alpha \to 1$. 

We highlight that a particular case of Eqs.~(\ref{Renyiln}) and~(\ref{generalizedRenyi}) occurs when $\Ln (x) = \ln_q (x)$, leading to biparametric objects in the Rényi-Tsallis context:
\begin{equation}
     \mathcal{S}_{\alpha, q}^{RT} (\rho) = \frac{1}{1-\alpha} \ln_q \left( \int_{-\infty}^\infty [\rho(x)]^{\alpha} dx \right)
\end{equation}%
and
\begin{equation}
\label{generalizedRenyilnq}
    \mathcal{D}_{\alpha, q}^{RT} (\rho, \rho_0) = \frac{1}{\alpha-1} \ln_q \left( \int_{-\infty}^\infty \frac{[\rho(x)]^{\alpha}}{[\rho_0(x)]^{\alpha-1}} dx  \right).
\end{equation}%
In a similar way, we can consider the Kaniadakis-motivated version of these two expressions by replacing $\ln_q(x)$ by $\ln_{\kappa}(x)$, that is, we have 
\begin{equation}
\mathcal{S}_{\alpha, \kappa}^{RK} (\rho) = \frac{1}{1-\alpha} \ln_{\kappa} \left( \int_{-\infty}^\infty [\rho(x)]^{\alpha} dx \right)
\end{equation}%
and
\begin{equation}
\label{generalizedRenyi-ln-kaniadakis}
\mathcal{D}_{\alpha, \kappa}^{RK} (\rho, \rho_0) = \frac{1}{\alpha-1} \ln_{\kappa} \left( \int_{-\infty}^\infty \frac{[\rho(x)]^{\alpha}}{[\rho_0(x)]^{\alpha-1}} dx  \right).
\end{equation}%

\subsection{Bregman divergence}

A third family of divergences was proposed by Bregman~\cite{bregmandivergence}. In this context, we consider
\begin{equation}
\label{bregman-phi}
    \mathcal{D}^B(\rho, \rho_0)= \int_{-\infty}^{\infty} \left[ \, \psi(\rho) - \psi (\rho_0) - (\rho-\rho_0) \, \psi' (\rho_0) \,  \right] \, dx \, ,
\end{equation}%
where $\psi$ is a convex function and $\psi'$ is its derivative. Note that $\mathcal{D}^B(\rho, \rho_0) \geq 0$ and $\mathcal{D}^B(\rho_0, \rho_0) = 0$ follow immediately from the convexity of $\psi$.

Applications of the Bregman divergence~(\ref{bregman-phi}) usually rely on specific choices for $\psi$. In order to exemplify specific cases of this divergence and connect it with our previous discussion on entropies and divergences, we consider $\psi = - \rho \Ln (1/\rho)$, leading to the form
\begin{equation}
    \label{bregman-Ln}
\mathcal{D}_{\Ln}^{B}(\rho,\rho_0)  = \int_{-\infty}^\infty \left[ \rho \Ln \left( \frac{1}{\rho_0} \right) - \rho \Ln \left( \frac{1}{\rho} \right) -  \frac{(\rho - \rho_0)}{\rho_0} \frac{1}{g(1/\rho_0)}
 \right]  dx \, .
\end{equation}%
In this context, a first example is to employ $\psi = -\, \rho \, \ln (1/\rho)$, which recovers the Kullback-Leibler divergence~(\ref{KLdiv}). On the other hand, when we replace the logarithm by the $q$-logarithm (i.e., using $\psi = -\, \rho \, \ln_q (1/\rho)$) we are directed to the Tsallis statistics. Thus, after renaming $\mathcal{D}_{\Ln}^B$ by $\mathcal{D}^{BT}_q$, we verify that this particular case of the Bregman divergence can be written as~\cite{naudts2004continuity, abe2005necessity}
\begin{equation}
\label{bregman-tsallis}
\mathcal{D}^{BT}_q (\rho, \rho_0) =  \int_{-\infty}^\infty \left[ - \rho \ln_q \left( \frac{1}{\rho} \right) + (1-q) \, \rho_0 \ln_q \left( \frac{1}{\rho_0} \right)  + \,  q \,  \rho \ln_q \left( \frac{1}{\rho_0} \right) \right] dx\, . 
\end{equation}%
Note that, despite divergences~(\ref{qdiv}),~(\ref{generalizedRenyilnq}) and~(\ref{bregman-tsallis}) converging to the Kullback-Leibler divergence in the limit case ($q \rightarrow 1$ and $\alpha \rightarrow 1$), they differ for general $q$ and $\alpha$ values.

A third example of the Bregman divergence can be addressed in the context of Kaniadakis statistics. For this purpose, we employ $\psi = -\, \rho \, \ln_{\kappa} (1/\rho)$ and rename $\mathcal{D}_{\Ln}^B$ by $\mathcal{D}_{\kappa}^{BK}$ to propose the divergence%
\begin{equation}
\label{bregman-kaniadakis}
\mathcal{D}_{\kappa}^{BK} (\rho, \rho_0) = \int_{-\infty}^{\infty} \left[ \vphantom{ \left( \frac{\rho_0^{\kappa} + \rho_0^{-\kappa}}{2} \right)  } \rho  \ln_{\kappa}(\rho) - \rho 
   \ln_{\kappa}(\rho_0)  - (\rho-\rho_0) \left( \frac{\rho_0^{\kappa} + \rho_0^{-\kappa}}{2} \right)  \right] dx\, .
\end{equation}%
Just as the divergences~(\ref{qdiv}),~(\ref{generalizedRenyilnq}) and~(\ref{bregman-tsallis}), the divergences~(\ref{divergencia-kaniadakis}),~(\ref{generalizedRenyi-ln-kaniadakis}) and~(\ref{bregman-kaniadakis}) are distinct for $\kappa\neq0$ and $\alpha \neq 1$; $\mathcal{D}_{\kappa}^{BK}$ reduces to the Kullback-Leibler divergence when $\kappa \to 0$.

\subsection{Burbea-Rao divergence}

Further generalized entropies and divergences can be considered. An example is the $(h,\phi)$-entropy, given by~\cite{pardo2018statistical}
\begin{equation}
    \mathcal{S}_{\phi}^{h}(\rho) = h \left( \int_{-\infty}^{\infty} \rho \, \phi(\rho) dx \right)  ,
\end{equation}%
where $\phi(x)$, defined for positive $x$, is concave (convex) and $h(x)$ is differentiable and
increasing (decreasing) for any $x$. Particular choices of $h$ and $\phi$ lead to entropies previously presented. If $h(x)=x$ and $\phi(x) = \ln(1/x)$, we obtain again the Shannon entropy~(\ref{shannonentropy}) [and, similarly, $\phi(x) = \Ln(1/x)$ leads to Eq.~(\ref{Sgeneralizado})]. For $h(x) = \ln(x)/(1-\alpha)$ and $\phi(x) = x^{\alpha-1} $, we recover the Rényi entropy~(\ref{Renyientropy}) [likewise, $h(x) = \Ln(x)/(1-\alpha)$ provides Eq.~(\ref{Renyiln})].

 In addition, we notice that  
\begin{equation}
\label{burbea}
    \mathcal{D}_{\phi}^{h}(\rho,\rho_0) = \mathcal{S}_{\phi}^{h} \left( \frac{\rho+\rho_0}{2} \right) - \frac{\mathcal{S}_{\phi}^{h}(\rho) + \mathcal{S}_{\phi}^{h}(\rho_0)}{2}
\end{equation}%
is its corresponding divergence~\cite{burbea1982convexity}. In fact, $\mathcal{D}_{\phi}^{h}(\rho,\rho_0)\geq 0$ because $\mathcal{S}_{\phi}^{h}$ is concave and $\mathcal{D}_{\phi}^{h}(\rho,\rho_0) = 0$ if $\rho=\rho_0$. Moreover, in contrast to the Kullback-Leibler divergence and others previously presented, it immediately follows that the Burbea-Rao divergence, $\mathcal{D}_{\phi}^{h}$, is symmetric, that is, $\mathcal{D}_{\phi}^{h}(\rho,\rho_0) = \mathcal{D}_{\phi}^{h}(\rho_0,\rho)$. Particularly, the use of $h(x)=x$ and $\phi(x) = \ln(1/x)$ in $\mathcal{D}_{\phi}^{h}$ leads to a symmetric divergence based on the Shannon entropy (and, consequently, on $\mathcal{D}^{KL}$), which is the well-known Jensen-Shannon divergence~\cite{jensenshannondiv}:
\begin{equation}
\label{27}
    \mathcal{D}^{JS}(\rho,\rho_0) = \frac{ \mathcal{D}^{KL}\left(\rho, \frac{\rho + \rho_0}{2}\right) + \mathcal{D}^{KL}\left(\rho_0, \frac{\rho + \rho_0}{2}\right)}{2} \, .
\end{equation}%
Other choices of $h$ and $\phi$ based on generalized logarithms also result in symmetric divergences that can be relevant, for instance, in the context of Tsallis~\cite{10.1117/1.2177638}. A further immediate example could be the Jensen-Kaniadakis divergence.

Thus far, we have considered entropies and used them to obtain divergences, which are usually not symmetric, $\mathcal{D}(\rho, \rho_0) \neq \mathcal{D}(\rho_0, \rho)$ (an exception is $\mathcal{D}_{\phi}^{h}$). However, we can obtain a symmetric divergence $\mathcal{D}^{S}$ directly starting from a given $\mathcal{D}$ via \begin{equation}\label{symmetrization}
    \mathcal{D}^{S}(\rho, \rho_0) = \frac{1}{2} \left[ \mathcal{D}(\rho, \rho_0) + \mathcal{D}(\rho_0, \rho)  \right] .
\end{equation}%
Accordingly, we now have seven different divergences in the Tsallis context: $\mathcal{D}_q^{T}$, $\mathcal{D}_q^{TS}$,  $\mathcal{D}_q^{RT}$, $\mathcal{D}_q^{RTS}$,
$\mathcal{D}_q^{BT}$, $\mathcal{D}_q^{BTS}$, and $\mathcal{D}_{q}^{BRT}$. Here, $\mathcal{D}_{q}^{BRT}$ is $\mathcal{D}_{\phi}^{h}$ with $h(x)= x$ and $\phi(x)=\ln_{q}(1/x)$ and the index $S$ refers to use of~(\ref{symmetrization}), for instance, $\mathcal{D}_q^{TS}$ indicates the symmetrization ~(\ref{symmetrization}) of $\mathcal{D}_q^{T}$. Similarly, there are seven divergences in the Kaniadakis scenario.

\section{Divergences as free energies}
\label{Divergences-as-free-energies}

Up to this point, based on generalizations of the Shannon entropy, we have called attention to some families of divergences. In this section, we propose a different route to obtain families of divergences. To motivate this procedure we begin by reviewing the connection between the Kullback-Leibler divergence and the Helmholtz free energy~\cite{friston2023free}. In this direction, we note that any distribution $\rho_0$ can be written in terms of a convenient potential energy $V$ as the canonical distribution
\begin{equation}
\label{stationarysolution}
    \rho_0 = \frac{1}{Z} e^{-\beta V},
\end{equation}%
where $Z$ is the partition function and $\beta > 0$ can be seen as an inverse temperature-like parameter. In this setting, the Kullback-Leibler divergence~(\ref{KLdiv}) can be rewritten as
\begin{equation}
\label{eq3}
    \mathcal{D}^{KL}(\rho, \rho_0) =  \beta F - \beta F_0 \, ,
\end{equation}%
where 
\begin{equation}
\label{funcional}
F= \int_{-\infty}^{\infty} \rho V dx - \frac{1}{\beta} \int_{-\infty}^{\infty} \rho \ln \left(\frac{1}{\rho} \right)  dx 
\end{equation}%
is a free energy-like functional and
$F_0= - \frac{1}{\beta}\ln (Z)$ is equal to $F$ when $\rho=\rho_0$, corresponding to the Helmholtz free energy. As there is no restriction on the choice of the $\beta$ value as well as the origin of the potential energy $V$, we can fix  $\beta$ and $Z$ as  equal to $1$. In this case, the Kullback-Leibler divergence reduces to $F$, that is, $\mathcal{D}^{KL}(\rho, \rho_0)=  F$. 

We note that, via an $H$-theorem~\cite{Risken}, the free energy-like functional $F$ is connected with the Fokker-Planck equation
\begin{equation}
\label{UsualFPequation}
    \frac{\partial \rho}{\partial t} = D \frac{\partial^2 \rho}{\partial x^2} - \frac{\partial }{\partial x} (A \rho ) \, ,
\end{equation}%
where $\rho=\rho(x,t)$, $D=1/\beta$ is a constant coefficient and $A=-dV/dx$ is a drift force. Indeed, the $H$-theorem for this equation reads as
\begin{equation}
    \frac{dF}{dt}\le 0 \, .
\end{equation}%
This inequality indicates that $F$ decreases toward a minimum, which corresponds to the equilibrium value $F_0$. In particular, this theorem implies that $F-F_0 \ge 0$ and, as a consequence of Eq.~(\ref{eq3}), we reproduce $\mathcal{D}^{KL}(\rho, \rho_0) \ge 0$ and $\mathcal{D}^{KL}(\rho_0, \rho_0) = 0$. In this context, the reference probability distribution $\rho_0(x)$ is the stationary solution of Eq.~(\ref{UsualFPequation}), i.e., $\lim_{t\to\infty}\rho(x,t)$. Throughout this work, $\rho_0$ is used as a reference probability distribution in divergences and as stationary solutions.

\subsection{General approach}

In the following, we seek other possible families of divergences by employing an $H$-theorem for a generalization of Eq.~(\ref{UsualFPequation}). This generalized Fokker-Planck equation is given by 
\begin{equation}
  \label{gFP}
  \frac{\partial \rho}{\partial t}=\frac{\partial}{\partial x} \left({\omega(\rho, x)\frac{\partial}{\partial x}\frac{\delta \mathcal{F}[\rho]}{\delta \rho}}\right),
\end{equation}%
in which $\rho=\rho(x,t)$ is non-negative with $x\in \mathbb{R}$ and $t\ge 0$, $\omega(\rho,x)$ is a non-negative function, and $\mathcal{F}$ is a generalized free energy-like functional ~\cite{frank2005nonlinear}. An example of this functional could be 
\begin{equation}
  \label{Ft}
  \mathcal{F}[\rho]= r \left(\int_{-\infty}^\infty \chi(\rho,x)\,dx \right) ,
\end{equation}%
where $\chi (\rho, x)$ is a continuously differentiable function and $r(y)$ is another differentiable function. By appropriately choosing the functions $r$, $\chi$ and $\omega$, Eq.~(\ref{gFP}) can encompass both linear and nonlinear Fokker-Planck equations as particular cases. For instance, it can be immediately verified that Eq.~(\ref{UsualFPequation}) is recovered when $\omega (\rho, x) =   D \rho$ and $\mathcal{F}[\rho]=F/D$, corresponding to $r(x)=x$ and $\chi(\rho,x) = -\rho \ln (\rho_0/\rho)$ in~(\ref{Ft}). To simplify nomenclature in the continuation of this work, we refer to Eq.~(\ref{gFP}) as Fokker-Planck equation and to $\mathcal{F}$ as free energy.

In this context, we begin by taking the derivative of Eq.~(\ref{Ft}) with respect to time
\begin{equation}
    \label{dFdt} 
  \frac{d \mathcal{F}[\rho]}{dt}=\int_{-\infty}^\infty\frac{\delta \mathcal{F}[\rho]}{\delta \rho}\frac{\partial \rho}{\partial t}\,dx\,.
\end{equation}%
Thus, after using Eq.~(\ref{gFP}) in Eq.~(\ref{dFdt}) and performing an  integration by parts, we obtain
\begin{eqnarray}
    \frac{d \mathcal{F}[\rho]}{dt}&=&\int_{-\infty}^\infty\frac{\delta \mathcal{F}[\rho]}{\delta \rho}\frac{\partial}{\partial x} \left( \omega(\rho,x)\frac{\partial}{\partial x}\frac{\delta \mathcal{F}[\rho]}{\delta \rho} \right)dx \nonumber \\
    &=& \left. \left( \omega(\rho,x)\frac{\delta \mathcal{F}[\rho]}{\delta \rho}\frac{\partial}{\partial x}\frac{\delta \mathcal{F}[\rho]}{\delta \rho} \right) \right|_{-\infty}^{\infty} \! \! - \! \int_{-\infty}^\infty \! \omega(\rho,x) \left| {\frac{\partial}{\partial x}\frac{\delta \mathcal{F}[\rho]}{\delta \rho}} \right|^2 \! \! dx.
\end{eqnarray}%
If we assume that the term $\left( \omega(\rho,x)\frac{\delta \mathcal{F}[\rho]}{\delta \rho}\frac{\partial}{\partial x}\frac{\delta \mathcal{F}[\rho]}{\delta \rho}\right)$ vanishes sufficiently fast as $|x| \to \infty$, for each $t \ge 0$, it follows that
\begin{equation}
  \label{Hthm}
  \frac{d\mathcal{F}[\rho]}{dt}=-\int_{-\infty}^\infty \omega(\rho, x) \left| \frac{\partial}{\partial x}\frac{\delta \mathcal{F}[\rho]}{\delta \rho}\right|^2\, dx\le 0\,.
\end{equation}%
Since the function $\omega (\rho ,x)$ is non-negative by definition, this inequality can be interpreted as an $H$-theorem associated with  Eq.~(\ref{gFP}). Therefore, if an $H$-theorem is verified for a large class of functionals $\mathcal{F}[\rho]$, particular cases automatically verify an $H$-theorem because of Eq.~(\ref{Hthm}).

An additional remark is that a variational structure can be associated with the Fokker–Planck equation~(\ref{gFP}) and its $H$-theorem~(\ref{Hthm}) in terms of the generalized 2-Wasserstein pseudo-distance, defined by~\cite{DiMarinoPortinaleRadici+2024+941+974}
\begin{equation}\label{39}
    W_{2,\omega}(\rho^{(0)}, \rho^{(1)})^{2} = \inf \left\{  \int_{0}^{1} \int_{\mathbb{R}^{d}} \omega(\rho) \left| v \right|^2 \, d^{d}x \, dt \right\},
\end{equation}
where $\omega(\rho)$ is called the mobility function, and the infimum is taken over all $\rho$ and $v$ satisfying
\begin{equation}\label{40}
    \frac{\partial \rho}{\partial t} + \nabla \cdot ( \omega(\rho) \, v ) = 0 \, ,
\end{equation}
with $\rho^{(0)}(x)= \rho(x, 0)$ and $ \rho^{(1)}(x)=  \rho(x, 1)$.
In simple terms, a gradient flow is an equation of the form~\cite{Otto31012001}
\begin{equation}
    \frac{\partial \rho}{\partial t}=-\Grad\mc F[\rho]\,,
\end{equation}
where $\Grad$ refers to a generalized gradient with respect to an underlying metric. If this metric comes from $W_{2,\omega}$, it can be verified that
\begin{equation}
    \Grad\mc F[\rho]=-\nabla\cdot\left(\omega(\rho)\nabla\frac{\delta\mc F}{\delta \rho}\right)\,.
\end{equation}
Therefore, Eq.~(\ref{gFP}) can be interpreted as the gradient flow of the free-energy functional $\mc F$ with respect to the 2-Wasserstein pseudo-distance $W_{2,m}$ in one dimension. In particular, this implies that the solutions of Eq.~(\ref{gFP}) are curves of steepest descent of the free-energy functional $\mc F$, which provides a stronger statement than the $H$-theorem. Further consequences of the gradient flow structure can be found in Refs.~\cite{LISINI2012814,Otto31012001,09c8ab87175a4811957ef21e53ab9188, DiMarinoPortinaleRadici+2024+941+974,CARRILLO20101273}.

Supposing an $H$-theorem for a generic functional $\mathcal{F}$ means that the functional decreases over time and tends toward its minimum $\mathcal{F}[\rho_0]$, corresponding to the stationary solution $\rho_0 (x)$. Thus, the quantity $\mathcal{F}[\rho, \rho_0] - \mathcal{F}[\rho_0, \rho_0]$ is non-negative, where it was employed $\mathcal{F}[\rho, \rho_0]$ ($\mathcal{F}[\rho_0, \rho_0]$) instead of $\mathcal{F}[\rho]$ ($\mathcal{F}[\rho_0]$) in order to emphasize the double dependence in $\rho$ and $\rho_0$. Moreover, $\mathcal{F}[\rho, \rho_0] - \mathcal{F}[\rho_0, \rho_0] = 0$ if and only if $\rho=\rho_0$. Hence, this difference can be seen as a divergence $\mathcal{D}(\rho, \rho_0)$ based on the free energy $\mathcal{F}$: 
\begin{equation}
\label{Divergenciageneralizada}
    \mathcal{D}(\rho, \rho_0) = \mathcal{F}[\rho, \rho_0] - \mathcal{F}[\rho_0, \rho_0]\, .
\end{equation}%
Consequently, by construction, this $\mathcal{D}(\rho, \rho_0)$ is a generalization of Eq.~(\ref{eq3}), that is, of the Kullback-Leibler divergence. For simplicity of notation, the analogue of $\beta$ was incorporated in $\mathcal{F}$ or $\mathcal{D}$. As $\rho$ and $\rho_0$ are arbitrary in a divergence, we can suppose, for instance, that $\rho$ is chosen as the initial condition of the Fokker-Planck equation and the stationary solution is selected in order to coincide with $\rho_0$. In some cases, $\mathcal{F}[\rho_0, \rho_0]$ is equal to zero and, for this reason, we have 
\begin{equation}
\label{divergencia-funcional}
    \mathcal{D}(\rho, \rho_0) = \mathcal{F}[\rho, \rho_0] \text{~~~~(if~~} \mathcal{F}[\rho_0, \rho_0] = 0 \text{)} \, ,
\end{equation}%
that is, the divergence is equal to the free energy when its minimum value is equal to zero. This possibility was illustrated in the case of $\mathcal{D}^{KL}$ after Eq.~(\ref{funcional}). Another example follows when $\mathcal{D}_{\Ln}(\rho, \rho_0)$ is directly identified with $\mathcal{F}[\rho, \rho_0]$.

\section{Applications}\label{Applications}

In the following, our unified approach connecting free energies, divergences and Fokker-Planck equations, presented in Sec.~\ref{Divergences-as-free-energies}, is illustrated by considering the divergences exhibited in Sec.~\ref{section1} as well as a number of free energies. Some of these applications recover previous studies, illustrating the scope of our proposal.

\subsection{Tsallis statistics}

In the context of Tsallis statistics, we employ a free energy similar to~(\ref{funcional}):
\begin{equation}
\label{funcionalFq}
    \mathcal{F}_q^{T}= \frac{1}{D_q}\int_{-\infty}^\infty \rho V(x) dx  -   \int_{-\infty}^\infty \rho \ln_q \left( \frac{1}{\rho} \right) dx \, ,
\end{equation}%
where $D_q$ is a positive constant. In comparison with Eq.~(\ref{funcional}), $\ln(x)$ was replaced by $\ln_q (x)$ and $D$ by $D_q$; $D_q$ coincides with $D$ in the limit $q\rightarrow 1$. Equation~(\ref{funcionalFq}) inserted in Eq.~(\ref{gFP}), with $\omega =  D_q\rho $, leads to the (nonlinear) Fokker-Planck equation
\begin{equation}
\label{FPmeiosporosos}
    \frac{\partial \rho}{\partial t} = D_q \frac{\partial^2 \rho^{q} }{\partial x^2}  - \frac{\partial}{\partial x} [A \rho] \, ,
\end{equation}%
where $A=-dV/dx$ is a drift force. This equation has the same structure as the porous media equation~\cite{refId0,Aronson1986} and was intoduced as a Fokker-Planck equation in the context of Tsallis statistics~\cite{plastino1995non}. 

For a stationary solution $\rho_0(x)$, we have $\partial \rho/\partial t = 0$ and, therefore, Eq.~(\ref{FPmeiosporosos}) simplifies to%
\begin{equation}
     \frac{d}{dx} \left(  D_q \frac{d \, \rho_0^q}{dx}  -A \rho_0 \right) =0 \,.
\end{equation}%
Assuming that the potential $V$ is confining, the term in parentheses can be taken as zero. In this case, we obtain 
\begin{equation}
    \rho_0 (x) = \left[ K_1 - \frac{(q-1)}{q} \frac{V}{D_q} \right]^{\frac{1}{q-1}}\,,
\end{equation}%
where $K_1$ is an integration constant to be fixed by the normalization of $\rho_0$. It is worth mentioning that, although the potential $V$ in Eqs.~(\ref{funcionalFq}) and~(\ref{stationarysolution}) is generic, its connection with the stationary solution $\rho_0$ is not identical, as it depends on the specific dynamics.

This solution can be rewritten in terms of the $q$-exponential as
\begin{equation}
\label{solucaoestacionariaq}
    \rho_0(x)=N \text{e}_{2-q}^{-\beta V} \,, 
\end{equation}%
where $\beta= 1/(qK_1D_q)$ and $N= K_1^{\frac{1}{q-1}}$. The $q$-exponential is defined by $\text{e}_{q'}^{x} = [1+(1-q')x]^{1/(1-q')}$ and is the inverse of the $q$-logarithm $    \ln_{q'} (x) = (x^{1-q'}-1)/(1-q')$, given in~(\ref{lnq}).

From Eq.~(\ref{solucaoestacionariaq}), the potential $V$ can be expressed in terms of $\rho_0$ as  
\begin{equation} \label{V-tsallis}
    V(x)=-\frac{1}{\beta} \ln_{2-q} \left( \frac{\rho_0}{N}\right) = \frac{1}{\beta} \ln_{q} \left( \frac{N}{\rho_0}\right).
\end{equation}%
The substitution of this potential into Eq.~(\ref{funcionalFq}) results in
\begin{equation}\label{funcional-tsallis}
    \mathcal{F}_q^{T}[\rho, \rho_0] = \int_{-\infty}^\infty  \rho \left[ -\ln_q \left( \frac{1}{\rho} \right) + q \ln_q \left( \frac{1}{\rho_0} \right) \right] dx  + \,  q \, K_1  \ln_q (N)  
\end{equation}%
and, consequently,
\begin{equation}\label{funcional-tsallis-rho_0}
\mathcal{F}_q^{T} [\rho_0, \rho_0] =  (q-1)
    \int_{-\infty}^\infty   \rho_0 \ln_q \left( \frac{1}{\rho_0} \right)  dx +q \, K_1 \ln_q (N) \,. 
\end{equation}%
Observe that the difference $\mathcal{F}_q^{T} [\rho, \rho_0] - \mathcal{F}_q^{T} [\rho_0, \rho_0]$ coincides with the divergence given in Eq.~(\ref{bregman-tsallis}), $\mathcal{D}^{BT}_q (\rho, \rho_0)$, illustrating a specific situation of the general relation~(\ref{Divergenciageneralizada}); $\mathcal{D}^{BT}_q$ was first considered in~\cite{abe2005necessity}.

\subsection{Overdamped system}

Before considering applications to other families of entropies and divergences, we review the thermostatistics of a one-dimensional overdamped system of pairwise interacting particles investigated in Ref.~\cite{PhysRevLett.105.260601} in light of the findings obtained so far. In our framework, a starting point is to establish the free energy $F^{OS}$ for the overdamped system, which is composed of three terms. The first two terms correspond to an external potential $V$ and the Shannon entropy $S$ as in Eq.~(\ref{funcional}). The last contribution in $F^{OS}$, in the continuum limit, comes from%
\begin{equation}
\label{FOS.ultimotermo}
    \frac{1}{2} \int_{-\infty}^{\infty}      \int_{-\infty}^{\infty} \rho(x) \rho(x') V_{2}(x-x')dx dx' \, ,
\end{equation}
where $V_{2}(x-x')$ is a pairwise energy. If $V_{2}(x-x')$ is approximated by a contact interaction, then $V_{2}(x-x') \propto \delta(x-x')$. Thus, the last term in $F^{OS}$ reduces to $D_2 \int_{-\infty}^{\infty} \rho^{2} dx$, where $D_2$ is a positive (negative) constant for a repulsive (attractive) interaction.

From the definition of the $q$-logarithm, it follows that $ \rho^{2}  = \rho - \rho \ln_2 \left( 1/\rho\right)$ and, therefore, $\int_{-\infty}^{\infty} \rho^{2} dx = 1 -S_2^{T} (\rho)$ after using the definition of $S_q^{T}$ and the normalization condition for $\rho$. Thus, the last term of $F^{OS}$, originally given in Eq.~(\ref{FOS.ultimotermo}), can be written as $-D_2\,S_2^{T}+D_{2}$. This last result reveals that there is an entropic contribution with $q=2$ for the overdamped system and that the presence of a contact pairwise interaction is formally equivalent to an entropic contribution. Consequently, the free energy for the overdamped system is 
\begin{eqnarray}\label{49}
    F^{OS} &=&   \int_{-\infty}^{\infty} \rho V(x) dx - D_1 \int_{-\infty}^{\infty} \rho \ln \left( \frac{1}{\rho}  \right)dx + D_2 \int_{-\infty}^{\infty} \rho^{2}   dx  \, \nonumber \\
&=& \int_{-\infty}^{\infty} \rho V(x) dx - D_1 S - D_2 S_2^{T}+D_{2} \, .
\end{eqnarray}
This result demonstrates that the entropic contribution in $F^{OS}$ contains two parts, $-D_1\, S$ and $ - D_2\, S_2^{T} $ and, therefore, $F^{OS}$ can be viewed as an interpolation of the two particular cases. If $ - D_2\, S_2^{T} $ is irrelevant, $F^{OS}$ simplifies to $F$ in Eq.~(\ref{funcional}). On the other hand, if $-D_1\, S$ is negligible, $F^{OS}$ reduces to $D_q\mathcal{F}_q^{T} $ with $q=2$, where $\mathcal{F}_q^{T}$ is given in Eq.~(\ref{funcionalFq}). Thus, $F^{OS}$ departs slightly from the other $\mathcal{F}$'s along this manuscript due to a multiplicative factor as $D_{q}$. 
 
The Fokker-Planck equation for this system is obtained by substituting Eq.~(\ref{49}) into Eq.~(\ref{gFP}), resulting in
\begin{equation}\label{50}
    \frac{\partial \rho}{\partial t} = \frac{\partial}{\partial x} \left\{ \omega \left[ \frac{dV}{dx} + \left( \frac{D_{1}}{\rho} + 2D_{2}  \right) \frac{\partial \rho}{\partial x}\right]  \right\}\, .
\end{equation}
As expected, if the pairwise interaction is not relevant, this equation reduces to the standard Fokker-Planck equation~(\ref{UsualFPequation}) when $\omega = \rho$ and $D_1=D$. If the usual difusive term is omitted, Eq.~(\ref{50}) recovers Eq.~(\ref{FPmeiosporosos}) with $\omega=\rho$ and $q=2$.

For the stationary solution $\rho_0$ of Eq.~(\ref{50}), the term inside the curly brackets is constant. This constant is equal to zero because $\rho_0(x) \rightarrow 0$ when $x\rightarrow \pm \infty$. Thus, Eq.~(\ref{50}) reduces to \begin{equation}\label{54}
   \frac{dV}{dx} = - \frac{D_{1}}{\rho_0}  \frac{d \rho_0}{d x}  - 2D_{2} \frac{d \rho_0}{d x} 
\end{equation}
since $\omega \neq 0$. The integration of this equation results in%
\begin{equation}\label{52}
    V(x) = - D_1 \ln (\rho_0) -2D_{2} \rho_0  + \Bar{K}\, ,
\end{equation}
with $\Bar{K}$ being a constant.

The use of Eq.~(\ref{52}) in Eq.~(\ref{49}) leads to 
\begin{equation}
    F^{OS}[\rho, \rho_0] = \int_{-\infty}^{\infty} \rho \left[ -2D_{2} \rho_0 + D_2 \rho - D_{1} \ln \left( \frac{\rho_0}{\rho}  \right)   \right] dx + \Bar{K}
\end{equation}
and
\begin{equation}
    F^{OS}[\rho_0, \rho_0]  = -D_{2} \int_{-\infty}^{\infty} \rho_{0}^{2} \, dx + \Bar{K} \, .
\end{equation}
The divergence associated with the functional $F^{OS}$ via Eq.~(\ref{Divergenciageneralizada}) is
\begin{eqnarray}\label{56}
    \mathcal{D}^{OS}(\rho, \rho_0) &=& -D_{1}\int_{-\infty}^{\infty} \rho \ln \left( \frac{\rho_0}{\rho}  \right) dx + D_{2} \int_{-\infty}^{\infty} (\rho - \rho_0)^{2} dx \nonumber \\
    &=& D_1 \mathcal{D}^{KL} + D_2 \mathcal{D}_{2}^{BT} \, ,
\end{eqnarray}
where $\mathcal{D}^{KL}$ is the Kullback-Leibler divergence [Eq.~(\ref{KLdiv})] and $\mathcal{D}_{2}^{BT}$ is given in Eq.~(\ref{bregman-tsallis}) with $q=2$. 

Equation~(\ref{56}) shows that the composition rule for the divergence $\mathcal{D}^{OS}$ is the same as for the contribution of entropies in Eq.~(\ref{49}). Thus, if $D_2=0$, the resulting divergence would be $D_1 \mathcal{D}^{KL}$. When $D_{1} = 0$, the ensuing divergence is $D_2 \mathcal{D}_{2}^{BT}$, that is, a Bregman divergence [Eq.~(\ref{bregman-phi})] with $\psi(\rho) = D_{2}\rho^{2}$. 

As a final remark regarding the overdamped system, we briefly discuss Eq.~(\ref{50}) with $\omega = \rho$ and in the absence of an external potential, $V = 0$. This purely diffusive situation was investigated in Ref.~\cite{PhysRevE.67.031104}, emphasizing that there are two limiting regimes. The first is the usual regime occurring when $D_2 = 0$, yielding $\sigma \sim t^{1/2}$, where $\sigma$ is the standard deviation. The second limiting regime arises when $D_1 = 0$, leading to anomalous diffusion (subdiffusion) with $\sigma \sim t^{1/3}$. The usual regime dominates at short times, the anomalous regime emerges at long times, and an median regime exists at intermediate times.

\subsection{Kaniadakis statistics}

Another application of our formalism relating free energy and divergence concerns the Kaniadakis statistics. In this case, the analogue of Eq.~(\ref{funcionalFq}) is the free energy 
\begin{equation}
\label{funcionalF-kaniadakis}
\mathcal{F}_{\kappa}^{K}= \frac{1}{D_{\kappa}}\int_{-\infty}^\infty \rho V(x) dx  -   \int_{-\infty}^\infty \rho \ln_\kappa \left( \frac{1}{\rho} \right) dx \, .
\end{equation}%
The substitution of this free energy into the Fokker-Planck equation~(\ref{gFP}) yields 
\begin{equation}
    \label{FP-bregman-kaniadakis}
    \frac{\partial \rho}{\partial t} = \frac{\partial}{\partial x} \left\{ \omega  \left[ -\frac{A}{D_{\kappa}} + \frac{\partial}{\partial x} \ln_{\kappa}(\rho) + \frac{\partial}{\partial x} \left( \frac{\rho^k+\rho^{-k}}{2} \right)
\right]  \right\}\,,
\end{equation}%
where $A=-dV/dx$.

In order to obtain the stationary solution of Eq.~(\ref{FP-bregman-kaniadakis}), we proceed similarly to what was done in the context of Tsallis statistics. Along these lines, if $\rho_0$ is the stationary solution of Eq.~(\ref{FP-bregman-kaniadakis}) (supposing $V$ confining and, for definiteness, $\omega = D_{\kappa} \rho$), then the expression inside curly brackets, as well as the term in square brackets, must vanish for  $\rho=\rho_0$. This procedure leads to%
\begin{equation}
    -\frac{A}{D_{\kappa}} + \frac{d}{d x} \ln_{\kappa}(\rho_0) + \frac{d}{d x} \left( \frac{\rho_0^k+\rho_0^{-k}}{2} \right) = 0
\end{equation}
and, after an integration over $x$, it yields
\begin{equation}\label{V-kaniadakis}
    \frac{V(x)}{D_{\kappa}} = - \ln_{\kappa}(\rho_0) - \frac{\rho_0^k+\rho_0^{-k}}{2} + K_2 \, ,
\end{equation}%
where $K_2$ is a constant. The replacement of  $V(x)$ in Eq.~(\ref{funcionalF-kaniadakis}) results in 
\begin{equation}
\mathcal{F}_{\kappa}^{K}[\rho,\rho_0] = \int_{-\infty}^{\infty} \rho \left[  \ln_{\kappa}(\rho) -  \ln_{\kappa}(\rho_0) -  \left( \frac{\rho_0^k+\rho_0^{-k}}{2} \right) \right] dx +  K_2
\end{equation}%
and, therefore, 
\begin{equation}\label{55}
    \mathcal{F}_{\kappa}^{K}[\rho_0,\rho_0] = - \int_{-\infty}^{\infty}  \rho_0 \left( \frac{\rho_0^k+\rho_0^{-k}}{2} \right)   dx + K_2 \, .
\end{equation}%
The use of these two free energies in Eq.~(\ref{Divergenciageneralizada}) leads to the divergence $\mathcal{D}_{\kappa}^{BK} (\rho, \rho_0) = \mathcal{F}_{\kappa}^{K}[\rho,\rho_0] - \mathcal{F}_{\kappa}^{K}[\rho_0,\rho_0]$, introduced in~(\ref{bregman-kaniadakis}). The (nonlinear) Fokker-Planck equation~(\ref{FP-bregman-kaniadakis}) and $\mathcal{D}_{\kappa}^{BK}$ can also be found in~\cite{scarfone2009lie, hirica2022lie}.

\subsection{Inhomogeneous case}

In this subsection, we generalize the two previous applications incorporating $\Ln(x)$ [Eq.~(\ref{Lngeneralizado})] and a non-constant ``diffusion'' coefficient $\overline{D}(x)$. In this  inhomogeneous case, the free energy that replaces $\mathcal{F}_q^{T}$ and $\mathcal{F}_{\kappa}^{K}$ is
\begin{equation}
\label{funcionalFLn}
    \mathcal{F}_{\Ln}= \int_{-\infty}^\infty  \rho \overline{V}(x) \, dx  -   \int_{-\infty}^\infty  \rho \Ln \left( \frac{1}{\rho} \right) dx \,,
\end{equation}%
where $\overline{V}(x)$ is an effective potential such that
\begin{equation}\label{v_barra}
    \frac{d\overline{V}(x)}{dx} = - \frac{A(x)}{\overline{D}(x)} \, .
\end{equation}%
The quantity $\overline{V}$, introduced in~\cite{sicuro2016nonlinear}, encompasses the effects of $\overline{D}(x)$ and $A(x)$ in a single entity; if $\overline{D}$ is a constant $D$, $\overline{V} = V/D$.

The use of $\mathcal{F}_{\Ln}$ in Eq.~(\ref{gFP}) yields the Fokker-Planck equation
\begin{equation}
\label{FP-Ln}
    \frac{\partial\rho}{\partial t} = \frac{\partial}{\partial x}    \left[- \omega \, \frac{A(x) }{\overline{D}(x)} + \frac{\omega}{\rho} \frac{\partial}{\partial x} \left( \frac{1}{g(1/\rho)} \right)  \right] .
\end{equation}%
We remark that if the usual drift term $-\partial(A\rho )/\partial x$ is intended, we need to employ $\omega(\rho,x)= \overline{D}(x) \rho$.

In the case of a stationary solution $\rho_0(x)$, by the same procedure as used for the Tsallis and Kaniadakis cases, Eq.~(\ref{FP-Ln}) turns into%
\begin{equation}
   \frac{d}{d x}    \left[ \omega \, \frac{ d\overline{V}}{dx} + \frac{\omega}{\rho_0} \frac{d}{d x} \left( \frac{1}{g(1/\rho_0)} \right)  \right] = 0 \, .
\end{equation}%
In a confining case ($\rho_0(x)\rightarrow 0$ when $x\rightarrow \pm \infty$), this equation implies that 
\begin{equation}
\label{passo1}
    \frac{d\overline{V}}{dx} = - \frac{1}{\rho_0}\frac{d}{dx}\left(\frac{1}{g(1/\rho_0)}\right) . 
\end{equation}%
By integrating both sides of this equation, applying integration by parts, and using the definition of $\Ln(x)$ [Eq.~(\ref{Lngeneralizado})], the right-hand side of Eq.~(\ref{passo1}) becomes
\begin{eqnarray}
    -\int \frac{1}{\rho_0} \frac{d}{dx} \left( \frac{1}{g(1/\rho_0)} \right) dx &=& \int \left( \frac{1}{g(1/\rho_0)} \right) \, d\!\left( \frac{1}{\rho_0} \right) - \frac{1}{\rho_0 \, g(1/\rho_0)} + K' \nonumber \\
    &=& \Ln \left( \frac{1}{\rho_0} \right) -\frac{1}{\rho_0 \, g(1/\rho_0)} + K \, .
\end{eqnarray}%
Thus, we verify that the effective potential $\overline{V}(x)$ is connected with the stationary solution $\rho_0$ via 
\begin{equation}
\label{V.funcionalFLn}
\overline{V}(x)= -\frac{1}{\rho_0 \, g(1/\rho_0)} + \Ln \left( \frac{1}{\rho_0} \right) + K \, .   
\end{equation}%
 In this expression, $K$ is an integration constant whose value can be found by employing the normalization condition after fixing the additive constant in $\overline{V}(x)$. Equations~(\ref{V.funcionalFLn}) and~(\ref{funcionalFLn}) lead to
\begin{equation}
    \mathcal{F}_{\Ln}[\rho,\rho_0] = - \int_{-\infty}^{\infty} \frac{\rho}{\rho_0 \, g(1/\rho_0)} dx + \int_{-\infty}^{\infty} \rho \Ln \left( \frac{1}{\rho_0} \right) dx   - \int_{-\infty}^{\infty} \rho \Ln \left( \frac{1}{\rho} \right) dx + K 
\end{equation}%
and, as consequence, 
\begin{equation}
    \mathcal{F}_{\Ln}[\rho_0, \rho_0]= - \int_{-\infty}^{\infty} \frac{1}{g(1/\rho_0)} dx  + K \, .
\end{equation}%
Thus, from Eq.~(\ref{Divergenciageneralizada}), we arrive again at the divergence~(\ref{bregman-Ln}), $\mathcal{D}_{\Ln}^{B}(\rho,\rho_0) = \mathcal{F}_{\Ln}[\rho, \rho_0] - \mathcal{F}_{\Ln}[\rho_0, \rho_0]$, that encompasses $\mathcal{D}_{q}^{T}$ and $\mathcal{D}_{\kappa}^{K}$ as particular cases. This result also demonstrates that the functional form of $\mathcal{D}_{\Ln}^{B}$ (when written in terms of $\rho$ and $\rho_0$) is not sensitive to $x$-dependence in $\overline{D}$.

\subsection{\texorpdfstring{$f$-divergence as a free energy}{f-Divergence as a free energy}}

The three previous applications were motivated by the form of Eq.~(\ref{funcional}), where we started from free energies that explicitly depend on $V$ to obtain divergences and Fokker-Planck equations. In the following, we consider divergences as free energies and the corresponding Fokker-Planck equations. These cases, in contrast, contain free energies that depend only on $\rho$ and $\rho_0$. In this direction, we first employ $\mathcal{D}_{\Ln}$ given by Eq.~(\ref{Dgeneralizado}) [or, alternatively, $\mathcal{D}_f$ in Eq.~(\ref{f-divergencia})] as a free energy to be inserted into the Fokker-Planck equation~(\ref{gFP}), that is,
\begin{equation}
\label{FP-Ln-generalizado}
    \frac{\partial \rho }{\partial t} = \frac{\partial}{\partial x} \left[ \omega \, \frac{\rho_0}{\rho} \frac{\partial}{\partial x} \left(\frac{1}{ g (\rho_0 / \rho)  } \right) \right] .
\end{equation}%
Specializing this equation to the case $\omega = D \rho$ and $g(y)=y$, we obtain the Fokker-Planck equation
\begin{equation}
\label{FP-kullback-leibler}
    \frac{\partial \rho }{\partial t} = \frac{\partial}{\partial x} \left[  \frac{D}{\rho_0}   \left( \rho_0 \frac{\partial \rho}{\partial x} - \rho \frac{d \rho_0}{d x} \right) \right] ,
\end{equation}%
which reduces to the usual one~(\ref{UsualFPequation}) if we employ the relation~(\ref{stationarysolution}) between $\rho_0$ and $V$. Other particular case of Eq.~(\ref{FP-Ln-generalizado}) occurs when we make use of $\omega = D_{q} \rho$ and $g(y) = y^{q}$ (with $q>0$), leading to
\begin{equation}
\label{FP-tsallis}
   \frac{\partial \rho }{\partial t} = \frac{\partial}{\partial x} \left[  \frac{q \,  D_{q} \, \rho^{q-1}}{\rho_0^{q}}   \left( \rho_0 \frac{\partial \rho}{\partial x} - \rho \frac{d \rho_0}{d x} \right) \right] .
\end{equation}%
Even when using relation~(\ref{solucaoestacionariaq}) between $\rho_0$ and $V$, this equation does not coincide with the nonlinear Fokker-Planck equation~(\ref{FPmeiosporosos}) for $q \neq 1$. This discrepancy arises because the divergences $\mathcal{D}_q^{T}$, Eq.~(\ref{qdiv}), and 
$\mathcal{D}^{BT}_q$, Eq.~(\ref{bregman-tsallis}), are different when $q\neq 1$. However, in the limit $q\rightarrow 1$, both cases reduce to the Kullback-Leibler divergence. In a similar way, when $\omega = D_{\kappa} \rho$ and $g(y)=2y/(y^\kappa+y^{-\kappa})$ (with $|\kappa| \leq 1$), we arrive at another new equation:
\begin{equation}
\label{FP-kullback-leibler3}   
   \frac{\partial \rho }{\partial t} = \frac{\partial}{\partial x} \left\{  \frac{D_{\kappa}}{2\rho_0} \left[ (1-\kappa)\left( \frac{\rho_0}{\rho} \right)^{\kappa} + (1+\kappa)\left( \frac{\rho_0}{\rho} \right)^{-\kappa}  \right]  \left( \rho_0 \frac{\partial \rho}{\partial x} - \rho \frac{d \rho_0}{d x} \right)  \vphantom{ \left[ 
 (1+\kappa)\left( \frac{\rho_0}{\rho} \right)^{-\kappa}  \right] } \right\} .
\end{equation}%
This equation is different from Eq.~(\ref{FP-bregman-kaniadakis}) even after using relation~(\ref{V-kaniadakis}) for $\kappa \neq 0$, because the divergences $\mathcal{D}_{\kappa}^{K}$, Eq.~(\ref{divergencia-kaniadakis}), and 
$\mathcal{D}^{BK}_{\kappa}$, Eq.~(\ref{bregman-kaniadakis}), are different for $\kappa \neq 0$; for $\kappa \rightarrow 0$, we have $\mathcal{D}_{0}^{K} = \mathcal{D}^{BK}_{0} = \mathcal{D}^{KL}$.

A stationary solution of Eq.~(\ref{FP-Ln-generalizado}) obeys the equation
\begin{equation}
\label{63}
    \frac{d}{dx} \left[ \omega \, \frac{\rho_0}{\rho} \frac{d}{dx} \left(\frac{1}{ g (\rho_0 / \rho)  } \right) \right] = 0 \,
\end{equation}%
and, thus, the term in square brackets is a constant. This constant is admitted to be zero because it is assumed that $\rho_0$ tends to zero when $x \rightarrow \pm \infty$. Therefore, considering that $\omega \, \rho_0/\rho$ is non-zero, we arrive at
\begin{equation}
    \frac{d}{dx} \left(\frac{1}{ g (\rho_0 / \rho)  } \right) = 0 \,.
\end{equation}%
The last result implies that $g(\rho_0/\rho)$ is a constant and, hence, the ratio $\rho_0/\rho$ must also be constant. Since $\rho$ and $\rho_0$ are normalized, we verify that this constant is equal to $1$ and, therefore, $\rho_0$ is the stationary solution of Eq.~(\ref{FP-Ln-generalizado}). Consequently, equations such as~(\ref{FP-kullback-leibler}),~(\ref{FP-tsallis}) and~(\ref{FP-kullback-leibler3}) are different but have the same stationary solution $\rho_0$. However, observe that the Fokker-Planck equation~(\ref{FP-Ln-generalizado}) has different dynamics toward equilibrium depending on the functions $g$ and $\omega$ for a given $\rho_0$. 

Since Eq.~(\ref{FP-kullback-leibler}) is linear and corresponds to the usual Fokker-Planck equation, it can be considered the simplest case of a Fokker-Planck equation that depends on $\rho_0$ instead of $V$. If we desire a dependence on a potential energy $V$, a relationship between $\rho_0$ and $V$ must be incorporated as an extra condition, since it does not arise from Eq.~(\ref{FP-kullback-leibler}). From a phenomenological point of view, equations such as~(\ref{FP-Ln-generalizado}) may be useful in situations where only the stationary solution $\rho_0$ is known, or when there is no need to specify a potential energy $V$.

\subsection{Rényi divergence as a free energy}

In a second example of using a divergence as a free energy, we have $\mathcal{D}_{\alpha, \Ln}^{R} (\rho, \rho_0)$ [Eq.~(\ref{generalizedRenyi})] as the $\mathcal{F}$ to be inserted in Eq.~(\ref{gFP}), giving
\begin{equation}
\label{FP-Renyi-Ln-generalizado}
    \frac{\partial \rho }{\partial t} = \frac{\partial}{\partial x} \left[ \frac{\omega \, \alpha}{(\alpha-1)} \frac{1}{g(B)} \, \frac{\partial}{\partial x}  \left( \frac{\rho}{\rho_0} \right)^{\alpha-1}  \right]  ,
\end{equation}%
where $B = \int_{-\infty}^{+\infty}\frac{\rho(x,t)^{\alpha}}{\rho_0(x)^{\alpha-1}} dx$. Three examples of this equation are related to $g(y)=y$ [Eq.~(\ref{Renyidiv})], $g(y)=y^{q}$ (with $q>0$) [Eq.~(\ref{generalizedRenyilnq})], and $g(y) = 2y/(y^{\kappa} + y^{-\kappa})$ (with $|\kappa| \leq 1$) [Eq.~(\ref{generalizedRenyi-ln-kaniadakis})]. 

Despite Eq.~(\ref{FP-Renyi-Ln-generalizado}) being a nonlinear integro-differential partial equation, its stationary solution can be obtained. In fact, the stationary solution of Eq.~(\ref{FP-Renyi-Ln-generalizado}) obeys the equation 
\begin{equation}
    \frac{d}{dx} \left[ \frac{\omega \, \alpha}{(\alpha-1)} \frac{1}{g(B)} \, \frac{d}{d x}  \left( \frac{\rho}{\rho_0} \right)^{\alpha-1}  \right] = 0 \, ,
\end{equation}%
which implies that the term in braces is constant. This constant is assumed to be zero because we are supposing $\rho_0=0$ when $x \rightarrow \pm \infty$. Moreover, the use of $\omega \, \alpha/[(\alpha-1) \, g(B)]$ different from zero leads to
\begin{equation}
   \frac{d}{d x}  \left( \frac{\rho}{\rho_0} \right)^{\alpha-1}  = 0 \, 
\end{equation}%
and, therefore, the term in brackets is constant, which means that $\rho=\rho_0$. Similarly to the previous context of the $f$-divergence, Eq.~(\ref{FP-Renyi-Ln-generalizado}) exhibits different dynamics toward equilibrium governed by $g$ and $\alpha$, while maintaining the same equilibrium solution $\rho_0$.

Before concluding this subsection, we indicate a free energy in the Rényi context. Similarly to how the free energies~(\ref{funcionalFq}),~(\ref{funcionalF-kaniadakis}), and~(\ref{funcionalFLn}) were introduced, the entropies can be replaced by $\mathcal{S}_{\alpha, \Ln}^{R}$, Eq.~(\ref{Renyiln}). For the simplest case, we have the entropy $\mathcal{S}_{\alpha}^{R}$ [Eq.~(\ref{Renyientropy})], the free energy
\begin{equation}
\label{free-energy-renyi}
    \mathcal{F}_{\alpha}^{R}= \frac{1}{D_{\alpha}}\int_{-\infty}^\infty \rho V(x) dx  -  \frac{1}{1-\alpha} \ln \left( \int_{-\infty}^\infty \rho^{\alpha} dx \right) ,
\end{equation}%
and the Fokker-Planck equation
\begin{equation}
\label{}
    \frac{\partial\rho}{\partial t} = \frac{\partial}{\partial x}    \left[ \omega \left( - \frac{A}{D_{\alpha}} + \frac{\alpha}{Y} \rho^{\alpha-2} \frac{\partial \rho}{\partial x} \right)  \right] ,
\end{equation}%
where $Y = \int_{-\infty}^{\infty} \rho(x,t)^{\alpha} dx$. 
Supposing that $V$ is confining, the stationary solution of this equation leads to the term in parentheses being equal to zero and, after an integration over $x$, we obtain 
\begin{equation}
    \frac{V(x)}{D_{\alpha}} = \frac{\alpha}{(1-\alpha)Y} \, \rho_{0}^{\alpha-1} + K_{3} \, ,
\end{equation}%
with $K_{3}$ being a constant. Thus, the replacement of this $V$ in~(\ref{free-energy-renyi}) results in the free energy $\mathcal{F}_{\alpha}^{R}[\rho, \rho_0]$. Therefore, by using Eq.~(\ref{Divergenciageneralizada}), we obtain the divergence
\begin{eqnarray}
\tilde{
\mathcal{D}}_{\alpha}^{R}(\rho, \rho_0) &=&  \frac{1}{\alpha-1} \left[  \ln\left( \int_{-\infty}^{\infty}  \rho^{\alpha} \, dx \right)  - \ln\left( \int_{-\infty}^{\infty}  \rho_{0}^{\alpha}\, dx \right)  \right] \nonumber \\
&& - \frac{\alpha}{(\alpha-1)Y} \int_{-\infty}^{\infty}  \rho_{0}^{\alpha-1} (\rho - \rho_0) \, dx \, .
\end{eqnarray}%
Note that $
\mathcal{D}_{\alpha}^{R}$ and $\tilde{
\mathcal{D}}_{\alpha}^{R}$ are different for $\alpha \neq 1$; in the limit $\alpha \rightarrow 1$, $
\mathcal{D}_{\alpha}^{R}= \tilde{
\mathcal{D}}_{\alpha}^{R} = \mathcal{D}^{KL}$. Analogously, a more general divergence could be obtained replacing $\ln(x)$ with $\Ln(x)$ in Eq.~(\ref{free-energy-renyi}).

\subsection{Bregman divergence as a free energy}

In order to advance our discussion on Fokker-Planck equations and divergences as free energies, we now turn our attention to the Bregman framework. In this direction, we substitute Eq.~(\ref{bregman-Ln}) into Eq.~(\ref{gFP}), which leads to the following Fokker-Planck equation:
\begin{equation}
\label{fp-bregman-Ln}
    \frac{\partial\rho}{\partial t} = \frac{\partial}{\partial x}    \left[ \frac{\omega}{\rho} \frac{\partial}{\partial x} \left( \frac{1}{g(1/\rho)} \right) - \frac{\omega}{\rho_0} \frac{d}{d x} \left( \frac{1}{g(1/\rho_0)} \right)   \right] .
\end{equation}%
If $\rho=\rho_0$, the right-hand side of Eq.~(\ref{fp-bregman-Ln}) becomes identically zero, indicating that $\rho_0$ is the stationary solution. Instead of starting with Eq.~(\ref{bregman-Ln}), we could also employ Eq.~(\ref{bregman-phi}), that is, to consider $\mathcal{D}^{B}$ directly as a free energy. In this case, the Fokker-Planck equation~(\ref{gFP}) reduces to 
\begin{equation}
\label{74}
\frac{\partial \rho}{\partial t} = \frac{\partial}{\partial x} \left\{  \omega  \frac{\partial}{\partial x} \left[ \psi'(\rho) -  \psi'(\rho_0) \right]
  \right\} .
\end{equation}%
The stationary solution of this equation leads to $ \omega \, d [\psi'(\rho) -  \psi'(\rho_0) ] /d x= C$, where $C$ is a constant. If we employ $\rho(x) = 0$ when $x \rightarrow \pm \infty$, we obtain $C = 0$. In addition, taking advantage of the fact that $\omega \neq 0$, it follows that $\psi'(\rho) -  \psi'(\rho_0) = C_1$, where $C_1$ is another constant. Using again $\rho(\pm \infty) = 0$, it results that $C_1 = 0$ and, therefore, $\psi'(\rho) = \psi'(\rho_0)$. Since $\psi'(x)$ is an invertible function, it is derived that $\rho=\rho_0$.

As illustrations of Eq.~(\ref{fp-bregman-Ln}), we can consider $g(y)=y$, $g(y)=y^q$ (with $q> 0$) and $g(y)=2y/(y^\kappa+y^{-\kappa})$ (with $|\kappa|\leq 1$). In these three cases we obtain  the Fokker-Planck equations related to the usual statistical mechanics, the Tsallis framework and the Kaniadakis approach, respectively. However, in contrast to those cases that start with a free energy explicitly containing a potential energy $V$, these three equations are expressed in terms of $\rho$ and $\rho_0$. In this scenario, additional information is required to establish a relationship between $\rho_0$ and $V$. For instance, when $\omega = D \rho $ and $g(y) = y$, the Fokker-Planck equation~(\ref{FP-kullback-leibler}) is derived again and reduces to the usual form (Eq.~(\ref{UsualFPequation})) when Eq.~(\ref{stationarysolution}) is employed. Analogously, for $\omega = D_{q} \rho$ and $g(y) = y^{q}$, we verify that 
\begin{equation}
\label{}
   \frac{\partial \rho }{\partial t} = \frac{\partial}{\partial x} \left[ q\, D_{q} \, \rho   \left( \rho^{q-2} \frac{\partial \rho}{\partial x} - \rho_0^{q-2} \frac{d \rho_0}{d x} \right) \right] ,
\end{equation}%
which reduces to the (nonlinear) Fokker-Planck equation~(\ref{FPmeiosporosos}) when relation~(\ref{solucaoestacionariaq}) is employed. Similarly, when $\omega = D_{\kappa} \rho$ and $g(y)=2y/(y^\kappa+y^{-\kappa})$, we arrive at 
\begin{eqnarray}
\label{79}
   \frac{\partial \rho }{\partial t} &=& \frac{\partial}{\partial x} \left\{  \frac{D_{\kappa}}{2}   \left[ (1-\kappa) \rho^{-\kappa} + (1+\kappa) \rho^{\kappa} \right] \frac{\partial \rho}{\partial x} \right. \nonumber \\
    &&  \left. -\frac{D_{\kappa} \, \rho}{2 \rho_0} \left[ (1-\kappa) \rho_{0}^{-\kappa} + (1+\kappa) \rho_{0}^{\kappa} \right] \frac{d \rho_0}{d x} 
   \right\} ,
\end{eqnarray}%
which becomes the Fokker-Planck equation~(\ref{FP-bregman-kaniadakis}) after using the relation~(\ref{V-kaniadakis}).

We note that the last term within the brackets of Eq.~(\ref{fp-bregman-Ln}) leads to a usual drift term ($-\partial (A\rho)/\partial x$) only when $\omega(\rho, x ) \propto \rho$, that is, $\omega(\rho, x ) = \overline{D}(x) \rho$. In fact, by setting $-A / \rho$ equal to the last term in brackets of Eq.~(\ref{79}), we obtain
\begin{equation}
    A =  \frac{ \overline{D}}{\rho_0} \frac{d}{d x} \left( \frac{1}{g(1/\rho_0)} \right) .
\end{equation}%
As we can see, this result is in accordance with Eqs.~(\ref{v_barra}) and~(\ref{passo1}) and indicates that interpreting $\mathcal{D}_{\Ln}^{B}$ as a free energy is consistent with the usual drift term. Alternatively, if we start with Eq.~(\ref{74}) and $\omega = \overline{D}(x) \rho$, we arrive at $A=\overline{D}(x) \psi''(\rho_0) \, d\rho_0/dx$. However, in the contexts of the divergences $\mathcal{D}_{\Ln}$ and $\mathcal{D}_{\alpha, \Ln}^{R}$, a drift term $-\partial(A \rho)/\partial x$ cannot be obtained from the Fokker-Planck equations~(\ref{FP-Ln-generalizado}) and~(\ref{FP-Renyi-Ln-generalizado}) for a general function $g$. 

Despite $\mathcal{D}_{\Ln}$ [Eq.~(\ref{Dgeneralizado})] and $\mathcal{D}^{B}$ [Eq.~(\ref{bregman-phi})] being different (except for some particular cases), we call attention to the fact that $\mathcal{D}_{\Ln}$ can be viewed as a generalized Bregman divergence. For the generalized Bregman divergence $\mathcal{D}^{GB}$ employed here, we replace $\psi(\rho)$ by $\psi(\rho, \rho_0)$ in Eq.~(\ref{bregman-phi}), resulting in 
\begin{equation}
\label{bregman-extended}
    \mathcal{D}^{GB}(\rho, \rho_0) = \int_{-\infty}^{\infty} \left[ \, \psi(\rho, \rho_0) - \psi (\rho_0, \rho_0)  - (\rho-\rho_0) \, \psi' (\rho_0, \rho_0) \,  \right] \, dx \, ,
\end{equation}%
where, given $\rho_0$, $\psi (\rho, \rho_0)$ is convex with respect to $\rho$ and $\psi'(x, x_0) = \partial \psi(x, x_0)/\partial x$. Thus, due to the convexity of $\psi (\rho, \rho_0)$, we reach the inequality $\mathcal{D}^{GB}(\rho, \rho_0) \geq 0$ with $\mathcal{D}^{GB}(\rho_0, \rho_0) = 0$. Note that the integrand of $\mathcal{D}_{\Ln}$, $\psi(\rho, \rho_0) = -\rho \Ln (\rho_0/\rho)$, is a convex function because $\partial^2 \psi/\partial \rho^2 > 0$. In turn, the use of this $\psi(\rho, \rho_0)$ in~(\ref{bregman-extended}) leads to the announced result: $\mathcal{D}^{GB} =  \mathcal{D}_{\Ln}$. Considering that the connection between $\mathcal{D}_f$ and $\mathcal{D}_{\Ln}$ is given by $f(x) = -C x \Ln (1/x)$ (with $C$ being a positive constant), it follows that $\mathcal{D}^{GB} = \mathcal{D}_{f}$. These results show that an attempt to obtain a new divergence starting from $\mathcal{D}_{f}$ via the generalized Bregman divergence fails to yield a new outcome; this procedure merely reproduces $\mathcal{D}_{f}$. In addition, we observe that a generalized Bregman divergence can also be related to a free energy via Eq.~(\ref{divergencia-funcional}) and the corresponding Fokker-Planck equation~(\ref{FP-Ln-generalizado}) has $\rho_0$ as its stationary solution.

\subsection{Burbea-Rao divergence as a free energy}

We consider the divergence $\mathcal{D}_{\phi}^{h}(\rho,\rho_0)$, Eq.~(\ref{burbea}), as a free energy to be inserted into the Fokker-Planck equation~(\ref{gFP}). As a consequence, we obtain the equation
\begin{eqnarray}
\label{71}
    \frac{\partial \rho}{\partial t} &=& \frac{\partial}{\partial x} \left\{ \frac{\omega}{2} \frac{\partial}{\partial x} \left[ h'\left(\int_{-\infty}^{\infty} \left( \frac{\rho+\rho_0}{2} \right) \, \phi \left(\frac{\rho+\rho_0}{2} \right)  dx \right) \right. \right. 
 \nonumber \\
 && \times \left( \left( \frac{\rho+\rho_0}{2} \right) \, \phi'\left( \frac{\rho+\rho_0}{2} \right) + \phi \left( \frac{\rho+\rho_0}{2} \right) \right)   \nonumber \\
    &&  - \left. \left. h'\left(\int_{-\infty}^{\infty} \rho \, \phi (\rho) \, dx \right) \left[ \rho \, \phi'(\rho) + \phi(\rho) \right] \right] \right\}  .
\end{eqnarray}

As done for the previous Fokker-Planck equations, we first set $\partial \rho/\partial t = 0$ to obtain the stationary solution. Consequently, this time-independent solution of Eq.~(\ref{71}) may be obtained by taking the outer $x$-derivative to be zero, that is, $d\{\cdots\}/dx = 0$. Therefore, consistent with the fact that $\rho(x)\rightarrow 0$ when $x\rightarrow \pm \infty$, we have $\{\cdots\}= 0$ and, hence, $d[ \cdots ]/dx = 0$. Again, using the fact that $\rho(x)\rightarrow 0$ when $x\rightarrow \pm \infty$, we arrive at $[\cdots]=0$. In turn, a direct inspection shows that $\rho=\rho_0$ identically conduces to $[\cdots]=0$, indicating that $\rho_0$ is the stationary solution of Eq.~(\ref{71}).

Some divergences are symmetric in $\rho$ and $\rho_0$, while others are not. To highlight how this symmetry or its absence manifests in Fokker-Planck equations, we consider the Kullback-Leibler divergence $\mathcal{D}^{KL}$~(\ref{KLdiv}), the Jensen-Shannon one $\mathcal{D}^{JS}$~(\ref{27}), and the symmetrized Kullback-Leibler one $\mathcal{D}^{KLS}$, defined via Eq.~(\ref{symmetrization}). For $\mathcal{D}^{KL}$ as a free energy, Eq.~(\ref{FP-Ln-generalizado}) with $g(y)=y$, we obtain 
\begin{equation}
\label{FP-kullback-leibler-outra}
    \frac{\partial \rho }{\partial t} = \frac{\partial}{\partial x} \left[  \frac{ \, \, \omega^{KL}}{\rho \rho_0}   \left( \rho_0 \frac{\partial \rho}{\partial x} - \rho \frac{d \rho_0}{d x} \right) \right] ,
\end{equation}%
where $\omega^{KL}$ stands for the weight $\omega$. If $\omega^{KL} = D \rho$, this equation becomes Eq.~(\ref{FP-kullback-leibler}), which simplifies to the usual Fokker-Planck equation~(\ref{UsualFPequation}) when Eq.~(\ref{stationarysolution}) is used to express $\rho_0$ in terms of $V$. On the other hand, we verified that a particular case of $\mathcal{D}_{\phi}^{h}(\rho,\rho_0)$ is the Jensen-Shannon divergence $\mathcal{D}^{JS}$, Eq.~(\ref{27}). In this case, Eq.~(\ref{71}) reduces to 
\begin{equation}
\label{FP-jensen-shannon}
    \frac{\partial \rho }{\partial t} = \frac{\partial}{\partial x} \left[  \frac{\omega^{JS}}{2\rho (\rho + \rho_0)}   \left( \rho_0 \frac{\partial \rho}{\partial x} - \rho \frac{d \rho_0}{d x} \right) \right]  
\end{equation}%
when $h(x)=x$ and $\phi(x)= \ln (1/x)$, and $\omega = \omega^{JS}$. In addition, the Fokker-Planck equation~(\ref{gFP}) that corresponds to $\mathcal{D}^{KLS}$ used as a free energy is given by
\begin{equation}
\label{FP-divKLsymmetrical}
    \frac{\partial \rho }{\partial t} = \frac{\partial}{\partial x} \left[  \frac{ \, \, \omega^{KLS} (\rho+\rho_0)}{2\rho^2 \rho_0}   \left( \rho_0 \frac{\partial \rho}{\partial x} - \rho \frac{d \rho_0}{d x} \right) \right] , 
\end{equation}%
where $\omega = \omega^{KLS}$. 

Although these last three Fokker–Planck equations share the same stationary solution, $\rho = \rho_0$, they exhibit distinct dynamics when $\omega^{KL} = \omega^{JS} = \omega^{KLS}$. However, if we wish to
reduce Eqs.~(\ref{FP-kullback-leibler}),~(\ref{FP-jensen-shannon}) and~(\ref{FP-divKLsymmetrical})  to the usual Fokker-Planck equation, we need
to employ
\begin{equation}\label{83}
    \omega^{JS} = \frac{2 (\rho + \rho_0)}{\rho_0} \, \omega^{KL} 
\end{equation}%
and
\begin{equation}\label{84}
    \omega^{KLS} = \frac{2\rho}{\rho + \rho_0} \, \omega^{KL} \, ,
\end{equation}%
with $\omega^{KL} = D\rho$. These relationships illustrate the fact that the flexibility in choosing $\omega$ can be employed to obtain suitable properties for the Fokker-Planck equations that emerge from the use of divergences as free energies. Equations~(\ref{83}) and~(\ref{84}) also suggest that $\omega$ may include an additional dependence on $\rho_0$, allowing us to fine-tune the structure of the Fokker-Planck equation.

Before closing our discussion of the Burbea-Rao divergence, we call attention to the free energy 
\begin{equation}
\label{}
\mathcal{F}^{BR}= \frac{1}{D^{BR}}\int_{-\infty}^\infty \rho V(x) dx  -  h \left( \int_{-\infty}^{\infty} \rho \, \phi(\rho) dx \right) .
\end{equation}%
The last term on the right-hand side of $\mathcal{F}^{BR}$ includes, as particular cases, the other entropies considered in this work. Thus, $\mathcal{F}^{BR}$ gives us a generalization of the previous free energies. Following the reasoning developed in this work, $\mathcal{F}^{BR}$ can be used to construct the divergence 
\begin{equation}
\label{}
    \mathcal{D}^{BR}(\rho, \rho_0) = \mathcal{F}^{BR}[\rho, \rho_0] - \mathcal{F}^{BR}[\rho_0, \rho_0]\, ,
\end{equation}%
where $\rho_0$ is the stationary solution of the Fokker-Planck equation~(\ref{gFP}) with $\mathcal{F} \rightarrow \mathcal{F}^{BR}$ and $D^{BR}$ a ``diffusion" coefficient. We also observe that $\mathcal{D}^{BR}(\rho, \rho_0)$ is not symmetric in $\rho$ and $\rho_0$, in contrast to $\mathcal{D}_{\phi}^{h}(\rho, \rho_0)$.

Throughout this section, we considered several applications related to the divergences presented in Sec.~\ref{section1}. However, a much broader set of examples related to these divergences could be directly developed. For instance, within the Tsallis context, we pointed out the divergences $\mathcal{D}_q^{T}$, $\mathcal{D}_q^{TS}$,  $\mathcal{D}_q^{RT}$, $\mathcal{D}_q^{RTS}$,
$\mathcal{D}_q^{BT}$, $\mathcal{D}_q^{BTS}$, and $\mathcal{D}_{q}^{BRT}$, although we explored applications only for $\mathcal{D}_q^{T}$, $\mathcal{D}_q^{RT}$, and $\mathcal{D}_q^{BT}$.

\section{Conclusions}

As pointed out in the introduction, there are many efforts to investigate situations that exhibit significative deviations from those expected when using the formalism of the standard statistical mechanics. From a formal point of view, some useful theoretical models can be employed to address such discrepancies. A known example is the one based on the Tsallis entropy, where concepts like free energy and Fokker-Planck equation are omnipresent, and a connection between divergence and these objects has been observed. Another example is associated with the Kaniadakis entropy. Usually, it is desirable to have approaches sufficiently comprehensive to incorporate techniques and tools useful for different generalized statistical mechanics. In this broader picture, this work provides a unified approach that connects divergences, (generalized) free energies, (generalized) Fokker-Planck equations, and $H$-theorem.

Before addressing the applications of our unified approach, two remarks concerning the dynamics governed by the generalized Fokker-Planck equation~(\ref{gFP}) are in order. First, given the free energy $\mathcal{F}[\rho, \rho_0]$ [or, alternatively, the divergence $\mathcal{D}(\rho, \rho_0)$], the stationary solution is independent of the dynamics and equals $\rho_0$. On the other hand, the dynamics toward the equilibrium solution $\rho_0$ differ when distinct divergences $\mathcal{D}(\rho, \rho_0)$ and weight functions $\omega$ are considered.

Our unified approach was applied in several scenarios, emphasizing free energy as divergence and the corresponding Fokker-Planck equation together with its stationary solution. This formal apparatus was initially employed in the Tsallis and Kaniadakis frameworks, considering two different types of free energies (divergences) for each of them, thereby recovering results already available in the literature. These two contexts were incorporated by using the $q$-logarithm and the $\kappa$-logarithm, respectively, in place of the standard logarithm. Our formalism also enables us to obtain new Fokker-Planck equations by considering more general divergences. For instance, a further substitution of the logarithm was performed in order to encompass a broader context related to the $f$-divergence. In addition, a Rényi scenario was examined. Our findings were also illustrated within the Bregman framework. The last application of our method concerned the Burbea-Rao divergence. Thus, our investigation provides an integrated view of the subject and encompasses applications to illustrate our formalism, some of these are already known, while others are new.

In several of our applications, we started from a free energy $\mathcal{F}$ that explicitly contains a potential energy $V$. The minimum of $\mathcal{F}$ leads to a relation involving the stationary solution $\rho_0$ of the Fokker-Planck equation and $V$. In this context, the divergence between $\rho$ and $\rho_0$, $\mathcal{D}(\rho, \rho_0)$, is obtained as the difference between the free energy functional and its minimum value. This type of application of our formalism can be understood as arising immediately from a generalization of the standard free energy. The main thread of connections between free energies and Fokker-Planck equations found in the literature aligns with this particular application of our unified formalism.

In contrast, when we employ a divergence $\mathcal{D}(\rho, \rho_0)$ as a free energy $\mathcal{F}$, the resulting Fokker-Planck-like equation does not include an explicit dependence on the potential energy $V$, but presents an unconventional dependence on the stationary solution $\rho_0$. This family of Fokker-Planck equations is one of the main contributions of this work, as it can be useful in contexts where $V$ is unknown. In order for the free energy $\mathcal{F}$ to have an explicit dependence on $V$, an extra relation between $\rho_0$ and $V$ must be incorporated. For instance, this occurs for the Bregman divergence when we suppose that the Fokker-Planck equation contains a usual drift term $-\partial (A \rho) / \partial x$. For other divergences, the possibility of considering a usual drift term is not consistent in general. We emphasize that, from a phenomenological point of view, the use of a Fokker-Planck-like equation that depends only on $\rho_0$ could be a useful formalism for situations where no explicit potential energy $V$ is evident. Our findings suggest that this is a promising approach in scenarios where only the stationary behavior of the system is known, without any reference to a possible potential energy. In other words, the lack of an explicit potential energy $V$ may not be a drawback when investigating a system, but rather a less restrictive way of approaching it at equilibrium.

\section*{Acknowledgements}

The authors thank A. A. B. Pessa for a careful reading of the manuscript and for insightful comments. We also acknowledge CNPq and CAPES (Brazilian funding agencies) for their partial financial support.


\begin{thebibliography}{10}
	\expandafter\ifx\csname url\endcsname\relax
	\def\url#1{\texttt{#1}}\fi
	\expandafter\ifx\csname urlprefix\endcsname\relax\def\urlprefix{URL }\fi
	\expandafter\ifx\csname href\endcsname\relax
	\def\href#1#2{#2} \def\path#1{#1}\fi
	
	\bibitem{Tsallis1988}
	C.~Tsallis, Possible generalization of {B}oltzmann-{G}ibbs statistics, Journal
	of Statistical Physics 52~(1) (1988) 479--487.
	\newblock \href {https://doi.org/10.1007/BF01016429}
	{\path{doi:10.1007/BF01016429}}.
	
	\bibitem{kaniadakis2001non}
	G.~Kaniadakis, Non-linear kinetics underlying generalized statistics, Physica A
	296~(3) (2001) 405--425.
	\newblock \href {https://doi.org/10.1016/S0378-4371(01)00184-4}
	{\path{doi:10.1016/S0378-4371(01)00184-4}}.
	
	\bibitem{tirnakli2016standard}
	U.~Tirnakli, E.~P. Borges, \href{https://www.nature.com/articles/srep23644}{The
		standard map: {F}rom {B}oltzmann-{G}ibbs statistics to {T}sallis statistics},
	Scientific reports 6~(1) (2016) 23644.
	\newblock \href {https://doi.org/10.1038/srep23644}
	{\path{doi:10.1038/srep23644}}.
	\newline\urlprefix\url{https://www.nature.com/articles/srep23644}
	
	\bibitem{tariq2021tumor}
	A.~T. Jamal, A.~B. Ishak, S.~Abdel-Khalek,
	\href{https://doi.org/10.1007/s00521-020-05518-x}{Tumor edge detection in
		mammography images using quantum and machine learning approaches}, Neural
	Computing and Applications 33~(13) (2021) 7773--7784.
	\newblock \href {https://doi.org/10.1007/s00521-020-05518-x}
	{\path{doi:10.1007/s00521-020-05518-x}}.
	\newline\urlprefix\url{https://doi.org/10.1007/s00521-020-05518-x}
	
	\bibitem{gupta2022stimulated}
	N.~Gupta, S.~Kumar, S.~B. Bhardwaj,
	\href{https://doi.org/10.1007/s12596-021-00822-8}{Stimulated {R}aman
		scattering of self-focused elliptical $q$-gaussian laser beam in plasma with
		axial density ramp: {E}ffect of ponderomotive force}, Journal of Optics
	51~(4) (2022) 819--833.
	\newblock \href {https://doi.org/10.1007/s12596-021-00822-8}
	{\path{doi:10.1007/s12596-021-00822-8}}.
	\newline\urlprefix\url{https://doi.org/10.1007/s12596-021-00822-8}
	
	\bibitem{makke2024data}
	N.~Makke, S.~Chawla,
	\href{https://doi.org/10.1093/pnasnexus/pgae467}{Data-driven discovery of
		{T}sallis-like distribution using symbolic regression in high-energy
		physics}, PNAS Nexus 3~(11) (2024) 467.
	\newblock \href {https://doi.org/10.1093/pnasnexus/pgae467}
	{\path{doi:10.1093/pnasnexus/pgae467}}.
	\newline\urlprefix\url{https://doi.org/10.1093/pnasnexus/pgae467}
	
	\bibitem{PhysRevResearch.7.L012081}
	C.~Beck, C.~Tsallis,
	\href{https://link.aps.org/doi/10.1103/PhysRevResearch.7.L012081}{Anomalous
		velocity distributions in slow quantum-tunneling chemical reactions},
	Physical Review Research 7 (2025) L012081.
	\newblock \href {https://doi.org/10.1103/PhysRevResearch.7.L012081}
	{\path{doi:10.1103/PhysRevResearch.7.L012081}}.
	\newline\urlprefix\url{https://link.aps.org/doi/10.1103/PhysRevResearch.7.L012081}
	
	\bibitem{tsallis2009introduction}
	C.~Tsallis,
	\href{https://link.springer.com/book/10.1007/978-0-387-85359-8}{Introduction
		to nonextensive statistical mechanics: {A}pproaching a complex world},
	Springer, 2023.
	\newline\urlprefix\url{https://link.springer.com/book/10.1007/978-0-387-85359-8}
	
	\bibitem{drepanou2022kaniadakis}
	N.~Drepanou, A.~Lymperis, E.~N. Saridakis, K.~Yesmakhanova,
	\href{https://doi.org/10.1140/epjc/s10052-022-10415-9}{Kaniadakis holographic
		dark energy and cosmology}, The European Physical Journal C 82~(5) (2022)
	449.
	\newblock \href {https://doi.org/10.1140/epjc/s10052-022-10415-9}
	{\path{doi:10.1140/epjc/s10052-022-10415-9}}.
	\newline\urlprefix\url{https://doi.org/10.1140/epjc/s10052-022-10415-9}
	
	\bibitem{irshad2023effect}
	M.~Irshad, A.~U. Rahman, M.~Khalid, S.~Khan, B.~M. Alotaibi, L.~S. El-Sherif,
	S.~A. El-Tantawy,
	\href{https://pubs.aip.org/aip/pof/article/35/10/105116/2915698}{Effect of
		$\kappa$-deformed {K}aniadakis distribution on the modulational instability
		of electron-acoustic waves in a non-{M}axwellian plasma}, {Physics of Fluids}
	35~(10) (2023) 105116.
	\newblock \href {https://doi.org/10.1063/5.0171327}
	{\path{doi:10.1063/5.0171327}}.
	\newline\urlprefix\url{https://pubs.aip.org/aip/pof/article/35/10/105116/2915698}
	
	\bibitem{e25030478}
	A.~S. Martinez, W.~V. de~Abreu,
	\href{https://www.mdpi.com/1099-4300/25/3/478}{{T}he scientific contribution
		of the {K}aniadakis entropy to nuclear reactor physics: {A} brief review},
	Entropy 25~(3) (2023) 478.
	\newblock \href {https://doi.org/10.3390/e25030478}
	{\path{doi:10.3390/e25030478}}.
	\newline\urlprefix\url{https://www.mdpi.com/1099-4300/25/3/478}
	
	\bibitem{e26050406}
	G.~Kaniadakis, \href{https://www.mdpi.com/1099-4300/26/5/406}{Relativistic
		roots of $\kappa$-entropy}, Entropy 26~(5) (2024) 406.
	\newblock \href {https://doi.org/10.3390/e26050406}
	{\path{doi:10.3390/e26050406}}.
	\newline\urlprefix\url{https://www.mdpi.com/1099-4300/26/5/406}
	
	\bibitem{e27030247}
	D.~T. Hristopulos, S.~L. E.~F. da~Silva, A.~M. Scarfone,
	\href{https://www.mdpi.com/1099-4300/27/3/247}{{T}wenty years of {K}aniadakis
		entropy: {C}urrent trends and future perspectives}, Entropy 27~(3) (2025)
	247.
	\newblock \href {https://doi.org/10.3390/e27030247}
	{\path{doi:10.3390/e27030247}}.
	\newline\urlprefix\url{https://www.mdpi.com/1099-4300/27/3/247}
	
	\bibitem{LENZI2000337}
	E.~K. Lenzi, R.~S. Mendes, L.~R. {da Silva},
	\href{https://www.sciencedirect.com/science/article/pii/S0378437100000078}{Statistical
		mechanics based on {R}ényi entropy}, Physica A: Statistical Mechanics and
	its Applications 280~(3) (2000) 337--345.
	\newblock \href {https://doi.org/https://doi.org/10.1016/S0378-4371(00)00007-8}
	{\path{doi:https://doi.org/10.1016/S0378-4371(00)00007-8}}.
	\newline\urlprefix\url{https://www.sciencedirect.com/science/article/pii/S0378437100000078}
	
	\bibitem{plastino1995non}
	A.~R. Plastino, A.~Plastino,
	\href{https://www.sciencedirect.com/science/article/pii/0378437195002111}{Non-extensive
		statistical mechanics and generalized {F}okker-{P}lanck equation}, Physica A
	222~(1-4) (1995) 347--354.
	\newblock \href {https://doi.org/https://doi.org/10.1016/0378-4371(95)00211-1}
	{\path{doi:https://doi.org/10.1016/0378-4371(95)00211-1}}.
	\newline\urlprefix\url{https://www.sciencedirect.com/science/article/pii/0378437195002111}
	
	\bibitem{PhysRevE.54.R2197}
	C.~Tsallis, D.~J. Bukman,
	\href{https://link.aps.org/doi/10.1103/PhysRevE.54.R2197}{Anomalous diffusion
		in the presence of external forces: {E}xact time-dependent solutions and
		their thermostatistical basis}, Physical Review E 54 (1996) R2197--R2200.
	\newblock \href {https://doi.org/10.1103/PhysRevE.54.R2197}
	{\path{doi:10.1103/PhysRevE.54.R2197}}.
	\newline\urlprefix\url{https://link.aps.org/doi/10.1103/PhysRevE.54.R2197}
	
	\bibitem{hirica2022lie}
	I.-E. Hirica, C.-L. Pripoae, G.-T. Pripoae, V.~Preda, Lie symmetries of the
	nonlinear {F}okker-{P}lanck equation based on weighted {K}aniadakis entropy,
	Mathematics 10~(15) (2022) 2776.
	\newblock \href {https://doi.org/10.3390/math10152776}
	{\path{doi:10.3390/math10152776}}.
	
	\bibitem{shiino2001free}
	M.~Shiino, \href{https://doi.org/10.1063/1.1367327}{Free energies based on
		generalized entropies and {$H$}-theorems for nonlinear {F}okker-{P}lanck
		equations}, Journal of Mathematical Physics 42~(6) (2001) 2540--2553.
	\newline\urlprefix\url{https://doi.org/10.1063/1.1367327}
	
	\bibitem{kaniadakis2001h}
	G.~Kaniadakis,
	\href{https://www.sciencedirect.com/science/article/pii/S0375960101005436}{{$H$}-theorem
		and generalized entropies within the framework of nonlinear kinetics},
	Physics Letters A 288~(5-6) (2001) 283--291.
	\newblock \href {https://doi.org/https://doi.org/10.1016/S0375-9601(01)00543-6}
	{\path{doi:https://doi.org/10.1016/S0375-9601(01)00543-6}}.
	\newline\urlprefix\url{https://www.sciencedirect.com/science/article/pii/S0375960101005436}
	
	\bibitem{schwammle2007consequences}
	V.~Schw{\"a}mmle, F.~D. Nobre, E.~M.~F. Curado,
	\href{https://link.aps.org/doi/10.1103/PhysRevE.76.041123}{Consequences of
		the {$H$}-theorem from nonlinear {F}okker-{P}lanck equations}, Physical
	Review E 76~(4) (2007) 041123.
	\newblock \href {https://doi.org/10.1103/PhysRevE.76.041123}
	{\path{doi:10.1103/PhysRevE.76.041123}}.
	\newline\urlprefix\url{https://link.aps.org/doi/10.1103/PhysRevE.76.041123}
	
	\bibitem{schwammle2007general}
	V.~Schw{\"a}mmle, E.~M.~F. Curado, F.~D. Nobre,
	\href{https://doi.org/10.1140/epjb/e2007-00217-1}{A general nonlinear
		{F}okker-{P}lanck equation and its associated entropy}, The European Physical
	Journal B 58 (2007) 159--165.
	\newblock \href {https://doi.org/10.1140/epjb/e2007-00217-1}
	{\path{doi:10.1140/epjb/e2007-00217-1}}.
	\newline\urlprefix\url{https://doi.org/10.1140/epjb/e2007-00217-1}
	
	\bibitem{sicuro2016nonlinear}
	G.~Sicuro, P.~Rap{\v{c}}an, C.~Tsallis,
	\href{https://link.aps.org/doi/10.1103/PhysRevE.94.062117}{Nonlinear
		inhomogeneous {F}okker-{P}lanck equations: Entropy and free-energy time
		evolution}, Physical Review E 94~(6) (2016) 062117.
	\newblock \href {https://doi.org/10.1103/PhysRevE.94.062117}
	{\path{doi:10.1103/PhysRevE.94.062117}}.
	\newline\urlprefix\url{https://link.aps.org/doi/10.1103/PhysRevE.94.062117}
	
	\bibitem{jauregui2021stationary}
	M.~Jauregui, A.~L.~F. Lucchi, J.~H.~Y. Passos, R.~S. Mendes,
	\href{https://link.aps.org/doi/10.1103/PhysRevE.104.034130}{Stationary
		solution and {$H$}-theorem for a generalized {{F}okker-{P}lanck} equation},
	Physical Review E 104 (2021) 034130.
	\newblock \href {https://doi.org/10.1103/PhysRevE.104.034130}
	{\path{doi:10.1103/PhysRevE.104.034130}}.
	\newline\urlprefix\url{https://link.aps.org/doi/10.1103/PhysRevE.104.034130}
	
	\bibitem{SILVA200765}
	A.~T. Silva, E.~K. Lenzi, L.~R. Evangelista, M.~K. Lenzi, L.~R. {da Silva},
	\href{https://www.sciencedirect.com/science/article/pii/S0378437106009848}{Fractional
		nonlinear diffusion equation, solutions and anomalous diffusion}, Physica A:
	Statistical Mechanics and its Applications 375~(1) (2007) 65--71.
	\newblock \href {https://doi.org/https://doi.org/10.1016/j.physa.2006.09.001}
	{\path{doi:https://doi.org/10.1016/j.physa.2006.09.001}}.
	\newline\urlprefix\url{https://www.sciencedirect.com/science/article/pii/S0378437106009848}
	
	\bibitem{lenzi2024generalized}
	E.~K. Lenzi, M.~P. Rosseto, D.~W. Gryczak, L.~R. Evangelista, L.~R. da~Silva,
	M.~K. Lenzi, R.~S. Zola, Generalized kinetic equations with fractional
	time-derivative and nonlinear diffusion: {$H$}-theorem and entropy, Entropy
	26~(8) (2024) 673.
	\newblock \href {https://doi.org/10.3390/e26080673}
	{\path{doi:10.3390/e26080673}}.
	
	\bibitem{plastino2022h}
	A.~R. Plastino, R.~S. Wedemann, F.~D. Nobre, {$H$}-theorems for systems of
	coupled nonlinear {F}okker-{P}lanck equations, Europhysics Letters 139~(1)
	(2022) 11002.
	\newblock \href {https://doi.org/10.1209/0295-5075/ac7c30}
	{\path{doi:10.1209/0295-5075/ac7c30}}.
	
	\bibitem{evangelista2023nonlinear}
	L.~R. Evangelista, E.~K. Lenzi, Nonlinear {F}okker-{P}lanck equations,
	{$H$}-theorem and generalized entropy of a composed system, Entropy 25~(9)
	(2023) 1357.
	\newblock \href {https://doi.org/10.3390/e25091357}
	{\path{doi:10.3390/e25091357}}.
	
	\bibitem{plastino2024evolution}
	A.~R. Plastino, R.~S. Wedemann,
	\href{https://dx.doi.org/10.1088/1742-6596/2839/1/012017}{Evolution equations
		exhibiting {$H$}-theorems related to the {LMC} statistical measures of
		complexity}, Journal of Physics: Conference Series 2839~(1) (2024) 012017.
	\newblock \href {https://doi.org/10.1088/1742-6596/2839/1/012017}
	{\path{doi:10.1088/1742-6596/2839/1/012017}}.
	\newline\urlprefix\url{https://dx.doi.org/10.1088/1742-6596/2839/1/012017}
	
	\bibitem{frank2005nonlinear}
	T.~D. Frank, Nonlinear {F}okker-{P}lanck equations: {F}undamentals and
	applications, Springer Science \& Business Media, 2005.
	\newblock \href {https://doi.org/10.1007/b137680} {\path{doi:10.1007/b137680}}.
	
	\bibitem{plastino1997universality}
	A.~Plastino, A.~R. Plastino,
	\href{https://doi.org/10.1016/S0375-9601(96)00942-5}{On the universality of
		thermodynamics' {L}egendre transform structure}, Physics Letters A 226~(5)
	(1997) 257--263.
	\newline\urlprefix\url{https://doi.org/10.1016/S0375-9601(96)00942-5}
	
	\bibitem{mendes1997some}
	R.~S. Mendes,
	\href{https://www.sciencedirect.com/science/article/pii/S0378437197001751}{Some
		general relations in arbitrary thermostatistics}, Physica A 242~(1-2) (1997)
	299--308.
	\newblock \href {https://doi.org/https://doi.org/10.1016/S0378-4371(97)00175-1}
	{\path{doi:https://doi.org/10.1016/S0378-4371(97)00175-1}}.
	\newline\urlprefix\url{https://www.sciencedirect.com/science/article/pii/S0378437197001751}
	
	\bibitem{hanel2011comprehensive}
	R.~Hanel, S.~Thurner, A comprehensive classification of complex statistical
	systems and an axiomatic derivation of their entropy and distribution
	functions, Europhysics Letters 93~(2) (2011) 20006.
	\newblock \href {https://doi.org/10.1209/0295-5075/93/20006}
	{\path{doi:10.1209/0295-5075/93/20006}}.
	
	\bibitem{abu2019effects}
	H.~A.~A. Alfeilat, A.~B.~A. Hassanat, O.~Lasassmeh, A.~S. Tarawneh, M.~B.
	Alhasanat, H.~S.~E. Salman, V.~B.~S. Prasath, Effects of distance measure
	choice on $k$-nearest neighbor classifier performance: {A} review, Big Data
	7~(4) (2019) 221--248.
	\newblock \href {https://doi.org/10.1089/big.2018.0175}
	{\path{doi:10.1089/big.2018.0175}}.
	
	\bibitem{deza2009encyclopedia}
	M.~M. Deza, E.~Deza, Encyclopedia of distances, Springer, 2009.
	\newblock \href {https://doi.org/10.1007/978-3-642-00234-2_1}
	{\path{doi:10.1007/978-3-642-00234-2_1}}.
	
	\bibitem{KullbackLeibler1951}
	S.~Kullback, R.~A. Leibler, \href{http://www.jstor.org/stable/2236703}{On
		information and sufficiency}, The Annals of Mathematical Statistics 22 (1951)
	79.
	\newline\urlprefix\url{http://www.jstor.org/stable/2236703}
	
	\bibitem{sakji2010fast}
	S.~Sakji-Nsibi, A.~Benazza-Benyahia, Fast scalable retrieval of multispectral
	images with {K}ullback-{L}eibler divergence, in: IEEE International
	Conference on Image Processing, 2010, pp. 2333--2336.
	\newblock \href {https://doi.org/10.1109/ICIP.2010.5653932}
	{\path{doi:10.1109/ICIP.2010.5653932}}.
	
	\bibitem{cui2012statistical}
	S.~Cui, M.~Datcu, Statistical wavelet subband modeling for multi-temporal {SAR}
	change detection, IEEE Journal of Selected Topics in Applied Earth
	Observations and Remote Sensing 5~(4) (2012) 1095--1109.
	\newblock \href {https://doi.org/10.1109/JSTARS.2012.2200655}
	{\path{doi:10.1109/JSTARS.2012.2200655}}.
	
	\bibitem{qin2015region}
	X.~Qin, H.~Zou, S.~Zhou, K.~Ji, Region-based classification of {SAR} images
	using {K}ullback-{L}eibler distance between generalized gamma distributions,
	IEEE Geoscience and Remote Sensing Letters 12~(8) (2015) 1655--1659.
	\newblock \href {https://doi.org/10.1109/LGRS.2015.2418217}
	{\path{doi:10.1109/LGRS.2015.2418217}}.
	
	\bibitem{cui2016comparison}
	S.~Cui, Comparison of approximation methods to {K}ullback-{L}eibler divergence
	between gaussian mixture models for satellite image retrieval, Remote Sensing
	Letters 7~(7) (2016) 651--660.
	\newblock \href {https://doi.org/10.1080/2150704X.2016.1177241}
	{\path{doi:10.1080/2150704X.2016.1177241}}.
	
	\bibitem{dhillon2003divisive}
	I.~S. Dhillon, S.~Mallela, R.~Kumar,
	\href{https://dl.acm.org/doi/10.5555/944919.944973}{A divisive information
		theoretic feature clustering algorithm for text classification}, The Journal
	of Machine Learning Research 3 (2003) 1265--1287.
	\newline\urlprefix\url{https://dl.acm.org/doi/10.5555/944919.944973}
	
	\bibitem{huang2008similarity}
	A.~Huang,
	\href{https://citeseerx.ist.psu.edu/document?repid=rep1&type=pdf&doi=b29bf8f6900b4b0258397f73957eabd1bb977ef4}{Similarity
		measures for text document clustering}, in: Proceedings of the Sixth New
	Zealand Computer Science Research Student Conference, Vol.~4, 2008, pp.
	9--56.
	\newline\urlprefix\url{https://citeseerx.ist.psu.edu/document?repid=rep1&type=pdf&doi=b29bf8f6900b4b0258397f73957eabd1bb977ef4}
	
	\bibitem{sun2020ml}
	B.~Sun, T.~Alkhalifah,
	\href{https://pubs.geoscienceworld.org/seg/geophysics/article/85/6/R477/592129/ML-descent-An-optimization-algorithm-for-full}{{ML}-descent:
		{A}n optimization algorithm for full-waveform inversion using machine
		learning}, Geophysics 85~(6) (2020) R477--R492.
	\newline\urlprefix\url{https://pubs.geoscienceworld.org/seg/geophysics/article/85/6/R477/592129/ML-descent-An-optimization-algorithm-for-full}
	
	\bibitem{ponti2017decision}
	M.~Ponti, J.~Kittler, M.~Riva, T.~de~Campos, C.~Zor, A decision cognizant
	{K}ullback-{L}eibler divergence, Pattern Recognition 61 (2017) 470--478.
	\newblock \href {https://doi.org/https://doi.org/10.1016/j.patcog.2016.08.018}
	{\path{doi:https://doi.org/10.1016/j.patcog.2016.08.018}}.
	
	\bibitem{kobayashi2022optimistic}
	T.~Kobayashi, Optimistic reinforcement learning by forward {K}ullback-{L}eibler
	divergence optimization, Neural Networks 152 (2022) 169--180.
	\newblock \href {https://doi.org/https://doi.org/10.1016/j.neunet.2022.04.021}
	{\path{doi:https://doi.org/10.1016/j.neunet.2022.04.021}}.
	
	\bibitem{clim2018kullback}
	A.~Clim, R.~D. Zota, G.~Tinica, The {K}ullback-{L}eibler divergence used in
	machine learning algorithms for health care applications and hypertension
	prediction: {A} literature review, Procedia Computer Science 141 (2018)
	448--453.
	\newblock \href {https://doi.org/https://doi.org/10.1016/j.procs.2018.10.144}
	{\path{doi:https://doi.org/10.1016/j.procs.2018.10.144}}.
	
	\bibitem{Rényi1961measures}
	A.~R{\'e}nyi,
	\href{https://projecteuclid.org/ebook/Download?urlid=bsmsp/1200512181&isFullBook=false}{On
		measures of entropy and information}, in: Proceedings of the Fourth Berkeley
	Symposium on Mathematical Statistics and Probability, Berkeley, California,
	USA, 1961, pp. 547--562.
	\newline\urlprefix\url{https://projecteuclid.org/ebook/Download?urlid=bsmsp/1200512181&isFullBook=false}
	
	\bibitem{he2003generalized}
	Y.~He, A.~B. Hamza, H.~Krim, A generalized divergence measure for robust image
	registration, IEEE Transactions on Signal Processing 51~(5) (2003)
	1211--1220.
	\newblock \href {https://doi.org/10.1109/TSP.2003.810305}
	{\path{doi:10.1109/TSP.2003.810305}}.
	
	\bibitem{li2019certified}
	B.~Li, C.~Chen, W.~Wang, L.~Carin,
	\href{https://proceedings.neurips.cc/paper_files/paper/2019/file/335cd1b90bfa4ee70b39d08a4ae0cf2d-Paper.pdf}{Certified
		adversarial robustness with additive noise}, Advances in neural information
	processing systems 32 (2019).
	\newline\urlprefix\url{https://proceedings.neurips.cc/paper_files/paper/2019/file/335cd1b90bfa4ee70b39d08a4ae0cf2d-Paper.pdf}
	
	\bibitem{mironov2019r}
	I.~Mironov, K.~Talwar, L.~Zhang,
	\href{https://arxiv.org/abs/1908.10530}{R\'enyi differential privacy of the
		sampled gaussian mechanism} (2019).
	\newline\urlprefix\url{https://arxiv.org/abs/1908.10530}
	
	\bibitem{chowdhury2020Rényi}
	B.~G. Chowdhury, S.~Datta, J.~R. David, R{\'e}nyi divergences from {E}uclidean
	quenches, Journal of High Energy Physics 2020~(4) (2020) 1--48.
	\newblock \href {https://doi.org/10.1007/JHEP04(2020)094}
	{\path{doi:10.1007/JHEP04(2020)094}}.
	
	\bibitem{knoblauch2022optimization}
	J.~Knoblauch, J.~Jewson, T.~Damoulas,
	\href{http://jmlr.org/papers/v23/19-1047.html}{An optimization-centric view
		on {B}ayes' rule: Reviewing and generalizing variational inference}, Journal
	of Machine Learning Research 23~(132) (2022) 1--109.
	\newline\urlprefix\url{http://jmlr.org/papers/v23/19-1047.html}
	
	\bibitem{Tsallisqdiv}
	C.~Tsallis, Generalized entropy-based criterion for consistent testing,
	Physical Review E 58 (1998) 1442--1445.
	\newblock \href {https://doi.org/10.1103/PhysRevE.58.1442}
	{\path{doi:10.1103/PhysRevE.58.1442}}.
	
	\bibitem{abe2003nonadditive}
	S.~Abe, Nonadditive generalization of the quantum {K}ullback-{L}eibler
	divergence for measuring the degree of purification, Physical Review A 68~(3)
	(2003) 032302.
	\newblock \href {https://doi.org/10.1103/PhysRevA.68.032302}
	{\path{doi:10.1103/PhysRevA.68.032302}}.
	
	\bibitem{rastegin2016quantum}
	A.~E. Rastegin, Quantum-coherence quantifiers based on the {T}sallis relative
	$\alpha$ entropies, Physical Review A 93~(3) (2016) 032136.
	\newblock \href {https://doi.org/10.1103/PhysRevA.93.032136}
	{\path{doi:10.1103/PhysRevA.93.032136}}.
	
	\bibitem{wang2021variational}
	Z.~Wang, O.~So, J.~Gibson, B.~Vlahov, M.~S. Gandhi, G.-H. Liu, E.~A. Theodorou,
	\href{https://roboticsproceedings.org/rss17/p073.pdf}{Variational inference
		{MPC} using {T}sallis divergence}, Robotics: Science and Systems (2021).
	\newline\urlprefix\url{https://roboticsproceedings.org/rss17/p073.pdf}
	
	\bibitem{fu2022tsallis}
	C.~Fu, P.~Xu, Y.~Huo, S.~Li, X.~Cai, Tsallis divergence as strategy for
	radioactive source search and location, Nuclear Science and Engineering
	196~(9) (2022) 1114--1124.
	\newblock \href {https://doi.org/10.1080/00295639.2022.2052551}
	{\path{doi:10.1080/00295639.2022.2052551}}.
	
	\bibitem{leleux2024sparse}
	P.~Leleux, B.~Lebichot, G.~Guex, M.~Saerens,
	\href{https://www.sciencedirect.com/science/article/pii/S0950705124007391}{Sparse
		randomized policies for {M}arkov decision processes based on {T}sallis
		divergence regularization}, Knowledge-Based Systems (2024) 112105\href
	{https://doi.org/https://doi.org/10.1016/j.knosys.2024.112105}
	{\path{doi:https://doi.org/10.1016/j.knosys.2024.112105}}.
	\newline\urlprefix\url{https://www.sciencedirect.com/science/article/pii/S0950705124007391}
	
	\bibitem{kaniadakisdivergence}
	R.-C. Sfetcu, S.-C. Sfetcu, V.~Preda, On {T}sallis and {K}aniadakis
	divergences, Mathematical Physics, Analysis and Geometry 25~(1) (2022) 7.
	\newblock \href {https://doi.org/10.1007/s11040-022-09420-x}
	{\path{doi:10.1007/s11040-022-09420-x}}.
	
	\bibitem{csiszar1974information}
	I.~Csisz{\'a}r,
	\href{https://www.fuw.edu.pl/~kostecki/scans/csiszar1978.pdf}{Information
		measures: {A} critical survey}, in: Transactions of the Seventh {P}rague
	Conference on Information Theory, Statistical Decision Functions, Random
	Processes, 1974, pp. 73--86.
	\newline\urlprefix\url{https://www.fuw.edu.pl/~kostecki/scans/csiszar1978.pdf}
	
	\bibitem{bregmandivergence}
	L.~M. Bregman, The relaxation method of finding the common point of convex sets
	and its application to the solution of problems in convex programming, USSR
	Computational Mathematics and Mathematical Physics 7~(3) (1967) 200--217.
	\newblock \href {https://doi.org/https://doi.org/10.1016/0041-5553(67)90040-7}
	{\path{doi:https://doi.org/10.1016/0041-5553(67)90040-7}}.
	
	\bibitem{burbea1982convexity}
	J.~Burbea, C.~Rao, On the convexity of some divergence measures based on
	entropy functions, IEEE Transactions on Information Theory 28~(3) (1982)
	489--495.
	\newblock \href {https://doi.org/10.1109/TIT.1982.1056497}
	{\path{doi:10.1109/TIT.1982.1056497}}.
	
	\bibitem{shiino1998h}
	M.~Shiino,
	\href{https://journals.jps.jp/doi/abs/10.1143/JPSJ.67.3658}{{$H$}-theorem
		with generalized relative entropies and the {T}sallis statistics}, Journal of
	the Physical Society of Japan 67~(11) (1998) 3658--3660.
	\newline\urlprefix\url{https://journals.jps.jp/doi/abs/10.1143/JPSJ.67.3658}
	
	\bibitem{qian2001relative}
	H.~Qian, \href{https://link.aps.org/doi/10.1103/PhysRevE.63.042103}{{R}elative
		entropy: {F}ree energy associated with equilibrium fluctuations and
		nonequilibrium deviations}, Physical Review E 63~(4) (2001) 042103.
	\newblock \href {https://doi.org/10.1103/PhysRevE.63.042103}
	{\path{doi:10.1103/PhysRevE.63.042103}}.
	\newline\urlprefix\url{https://link.aps.org/doi/10.1103/PhysRevE.63.042103}
	
	\bibitem{abe2005necessity}
	S.~Abe, G.~B. Bagci, Necessity of $q$-expectation value in nonextensive
	statistical mechanics, Physical Review E 71~(1) (2005) 016139.
	\newblock \href {https://doi.org/10.1103/PhysRevE.71.016139}
	{\path{doi:10.1103/PhysRevE.71.016139}}.
	
	\bibitem{pennini2021free}
	F.~Pennini, A.~Plastino, J.~Ya{\~n}ez, G.~L. Ferri,
	\href{https://www.sciencedirect.com/science/article/pii/S0378437120308037}{Free
		energies divergences as statistical quantifiers}, Physica A 564 (2021)
	125505.
	\newblock \href {https://doi.org/https://doi.org/10.1016/j.physa.2020.125505}
	{\path{doi:https://doi.org/10.1016/j.physa.2020.125505}}.
	\newline\urlprefix\url{https://www.sciencedirect.com/science/article/pii/S0378437120308037}
	
	\bibitem{6773024}
	C.~E. Shannon, A mathematical theory of communication, The Bell System
	Technical Journal 27~(3) (1948) 379--423.
	\newblock \href {https://doi.org/10.1145/584091.584093}
	{\path{doi:10.1145/584091.584093}}.
	
	\bibitem{Durrett_2019}
	R.~Durrett, Probability: Theory and Examples, 5th Edition, Cambridge Series in
	Statistical and Probabilistic Mathematics, Cambridge University Press, 2019.
	
	\bibitem{kaniadakisentropy}
	G.~Kaniadakis, Non-linear kinetics underlying generalized statistics, Physica A
	296~(3-4) (2001) 405--425.
	\newblock \href {https://doi.org/https://doi.org/10.1016/S0378-4371(01)00184-4}
	{\path{doi:https://doi.org/10.1016/S0378-4371(01)00184-4}}.
	
	\bibitem{naudts2004generalized}
	J.~Naudts, Generalized thermostatistics based on deformed exponential and
	logarithmic functions, Physica A 340~(1-3) (2004) 32--40.
	\newblock \href {https://doi.org/https://doi.org/10.1016/j.physa.2004.03.074}
	{\path{doi:https://doi.org/10.1016/j.physa.2004.03.074}}.
	
	\bibitem{ali1966general}
	S.~M. Ali, S.~D. Silvey, A general class of coefficients of divergence of one
	distribution from another, Journal of the Royal Statistical Society: Series B
	(Methodological) 28~(1) (1966) 131--142.
	\newblock \href {https://doi.org/10.1111/j.2517-6161.1966.tb00626.x}
	{\path{doi:10.1111/j.2517-6161.1966.tb00626.x}}.
	
	\bibitem{csiszar1972class}
	I.~Csisz{\'a}r, \href{https://doi.org/10.1007/bf02018661}{A class of measures
		of informativity of observation channels}, Periodica Mathematica Hungarica
	2~(1-4) (1972) 191--213.
	\newline\urlprefix\url{https://doi.org/10.1007/bf02018661}
	
	\bibitem{liese2006divergences}
	F.~Liese, I.~Vajda, On divergences and informations in statistics and
	information theory, IEEE Transactions on Information Theory 52~(10) (2006)
	4394--4412.
	\newblock \href {https://doi.org/10.1109/TIT.2006.881731}
	{\path{doi:10.1109/TIT.2006.881731}}.
	
	\bibitem{naudts2004continuity}
	J.~Naudts, Continuity of a class of entropies and relative entropies, Reviews
	in Mathematical Physics 16~(06) (2004) 809--822.
	\newblock \href {https://doi.org/10.1142/S0129055X04002151}
	{\path{doi:10.1142/S0129055X04002151}}.
	
	\bibitem{pardo2018statistical}
	L.~Pardo, Statistical inference based on divergence measures, Chapman and
	Hall/CRC, 2018.
	\newblock \href {https://doi.org/https://doi.org/10.1201/9781420034813}
	{\path{doi:https://doi.org/10.1201/9781420034813}}.
	
	\bibitem{jensenshannondiv}
	J.~Lin, Divergence measures based on the {S}hannon entropy, IEEE Transactions
	on Information Theory 37~(1) (1991) 145--151.
	\newblock \href {https://doi.org/10.1109/18.61115}
	{\path{doi:10.1109/18.61115}}.
	
	\bibitem{10.1117/1.2177638}
	A.~B. Hamza, \href{https://doi.org/10.1117/1.2177638}{{Nonextensive
			information-theoretic measure for image edge detection}}, Journal of
	Electronic Imaging 15~(1) (2006) 013011.
	\newblock \href {https://doi.org/10.1117/1.2177638}
	{\path{doi:10.1117/1.2177638}}.
	\newline\urlprefix\url{https://doi.org/10.1117/1.2177638}
	
	\bibitem{friston2023free}
	K.~Friston, L.~da~Costa, N.~Sajid, C.~Heins, K.~Ueltzh{\"o}ffer, G.~A.
	Pavliotis, T.~Parr,
	\href{https://www.sciencedirect.com/science/article/pii/S037015732300203X}{The
		free energy principle made simpler but not too simple}, Physics Reports 1024
	(2023) 1--29.
	\newblock \href {https://doi.org/https://doi.org/10.1016/j.physrep.2023.07.001}
	{\path{doi:https://doi.org/10.1016/j.physrep.2023.07.001}}.
	\newline\urlprefix\url{https://www.sciencedirect.com/science/article/pii/S037015732300203X}
	
	\bibitem{Risken}
	H.~Risken, The {F}okker-{P}lanck equation: {M}ethods of solution and
	applications, 2nd Edition, Springer, Berlin, 1996.
	
	\bibitem{DiMarinoPortinaleRadici+2024+941+974}
	S.~D. Marino, L.~Portinale, E.~Radici,
	\href{https://doi.org/10.1515/acv-2022-0076}{Optimal transport with nonlinear
		mobilities: {A} deterministic particle approximation result}, Advances in
	Calculus of Variations 17~(3) (2024) 941--974.
	\newblock \href {https://doi.org/10.1515/acv-2022-0076}
	{\path{doi:10.1515/acv-2022-0076}}.
	\newline\urlprefix\url{https://doi.org/10.1515/acv-2022-0076}
	
	\bibitem{Otto31012001}
	F.~Otto, \href{https://doi.org/10.1081/PDE-100002243}{The geometry of
		dissipative evolution equations: The porous medium equation}, Communications
	in Partial Differential Equations 26~(1-2) (2001) 101--174.
	\newblock \href {http://arxiv.org/abs/https://doi.org/10.1081/PDE-100002243}
	{\path{arXiv:https://doi.org/10.1081/PDE-100002243}}, \href
	{https://doi.org/10.1081/PDE-100002243} {\path{doi:10.1081/PDE-100002243}}.
	\newline\urlprefix\url{https://doi.org/10.1081/PDE-100002243}
	
	\bibitem{LISINI2012814}
	S.~Lisini, D.~Matthes, G.~Savaré,
	\href{https://www.sciencedirect.com/science/article/pii/S0022039612001544}{Cahn–{H}illiard
		and thin film equations with nonlinear mobility as gradient flows in
		weighted-{W}asserstein metrics}, Journal of Differential Equations 253~(2)
	(2012) 814--850.
	\newblock \href {https://doi.org/https://doi.org/10.1016/j.jde.2012.04.004}
	{\path{doi:https://doi.org/10.1016/j.jde.2012.04.004}}.
	\newline\urlprefix\url{https://www.sciencedirect.com/science/article/pii/S0022039612001544}
	
	\bibitem{09c8ab87175a4811957ef21e53ab9188}
	A.~Mielke, D.~R.~M. Renger, M.~A. Peletier,
	\href{https://doi.org/10.1515/jnet-2015-0073}{A generalization of {O}nsager's
		reciprocity relations to gradient flows with nonlinear mobility}, Journal of
	Non-Equilibrium Thermodynamics 41~(2) (2016) 141--149.
	\newblock \href {https://doi.org/10.1515/jnet-2015-0073}
	{\path{doi:10.1515/jnet-2015-0073}}.
	\newline\urlprefix\url{https://doi.org/10.1515/jnet-2015-0073}
	
	\bibitem{CARRILLO20101273}
	J.~A. Carrillo, S.~Lisini, G.~Savaré, D.~Slepčev,
	\href{https://www.sciencedirect.com/science/article/pii/S0022123609004261}{Nonlinear
		mobility continuity equations and generalized displacement convexity},
	Journal of Functional Analysis 258~(4) (2010) 1273--1309.
	\newblock \href {https://doi.org/https://doi.org/10.1016/j.jfa.2009.10.016}
	{\path{doi:https://doi.org/10.1016/j.jfa.2009.10.016}}.
	\newline\urlprefix\url{https://www.sciencedirect.com/science/article/pii/S0022123609004261}
	
	\bibitem{refId0}
	H.~Spohn, Surface dynamics below the roughening transition, Journal de Physique
	I 3~(1) (1993) 69--81.
	\newblock \href {https://doi.org/10.1051/jp1:1993117}
	{\path{doi:10.1051/jp1:1993117}}.
	
	\bibitem{Aronson1986}
	D.~G. Aronson, The porous medium equation, in: Lecture Notes in Mathematics,
	Vol. 1224, Chelsea, Berlin, 1986, p.~1.
	\newblock \href {https://doi.org/10.1007/BFb0072687}
	{\path{doi:10.1007/BFb0072687}}.
	
	\bibitem{PhysRevLett.105.260601}
	J.~S. Andrade, G.~F.~T. da~Silva, A.~A. Moreira, F.~D. Nobre, E.~M.~F. Curado,
	\href{https://link.aps.org/doi/10.1103/PhysRevLett.105.260601}{Thermostatistics
		of overdamped motion of interacting particles}, Phys. Rev. Lett. 105 (2010)
	260601.
	\newblock \href {https://doi.org/10.1103/PhysRevLett.105.260601}
	{\path{doi:10.1103/PhysRevLett.105.260601}}.
	\newline\urlprefix\url{https://link.aps.org/doi/10.1103/PhysRevLett.105.260601}
	
	\bibitem{PhysRevE.67.031104}
	E.~K. Lenzi, R.~S. Mendes, C.~Tsallis,
	\href{https://link.aps.org/doi/10.1103/PhysRevE.67.031104}{Crossover in
		diffusion equation: Anomalous and normal behaviors}, Phys. Rev. E 67 (2003)
	031104.
	\newblock \href {https://doi.org/10.1103/PhysRevE.67.031104}
	{\path{doi:10.1103/PhysRevE.67.031104}}.
	\newline\urlprefix\url{https://link.aps.org/doi/10.1103/PhysRevE.67.031104}
	
	\bibitem{scarfone2009lie}
	A.~M. Scarfone, T.~Wada,
	\href{https://doi.org/10.1590/S0103-97332009000400024}{Lie symmetries and
		related group-invariant solutions of a nonlinear {F}okker-{P}lanck equation
		based on the {S}harma-{T}aneja-{M}ittal entropy}, Brazilian Journal of
	Physics 39 (2009) 475--482.
	\newblock \href {https://doi.org/10.1590/S0103-97332009000400024}
	{\path{doi:10.1590/S0103-97332009000400024}}.
	\newline\urlprefix\url{https://doi.org/10.1590/S0103-97332009000400024}
	
\end{thebibliography}
\end{document}